\documentclass[12pt, draftcls, onecolumn]{IEEEtran}
\makeatletter
\def\subsubsection{\@startsection{subsubsection}
                                 {3}
                                 {\z@}
                                 {0ex plus 0.1ex minus 0.1ex}
                                 {0ex}
                             {\normalfont\normalsize\bfseries}}
\makeatother
\usepackage[T1]{fontenc}
\usepackage{ulem}
\usepackage{amsmath}
\allowdisplaybreaks
\usepackage{hhline}
\usepackage{graphicx}
\usepackage{yfonts,color}
\usepackage{soul,xcolor}
\usepackage{makecell}
\usepackage{verbatim}
\usepackage{amsmath}
\allowdisplaybreaks
\usepackage{amssymb}
\usepackage{amsthm}
\usepackage{float}
\usepackage{bm}
\usepackage{url}
\usepackage{array}
\usepackage{cite}
\usepackage{tikz}
\usepackage{enumitem}
\usepackage{framed}
\usepackage{balance}
\usepackage{epsfig,epstopdf}
\usepackage{booktabs}
\usepackage{courier}
\usepackage{caption}
\usepackage{subcaption}
\usepackage{algorithm}
\usepackage[noend]{algpseudocode}
\usepackage{csquotes}

\newcommand{\rom}[1]{\uppercase\expandafter{\romannumeral #1\relax}}
\usepackage{color}
\usepackage{soul,xcolor}

\newcommand{\secref}[1]{{Section~\ref{#1}}}
\newcommand{\figref}[1]{{Fig.~\ref{#1}}}
\newcommand\hlt[1]{\textcolor{black}{#1}}

\DeclareMathOperator*{\argmax}{arg\,max}

\usepackage{cancel}

\title{\hlt{Learning-based Spectrum Sensing and Access in Cognitive Radios via Approximate POMDPs}}
\author{Bharath Keshavamurthy and Nicol\`{o} Michelusi
\thanks{Part of this research appeared in the proceedings of IEEE ICC 2021 \cite{ICC:paper}.}
\thanks{This research has been funded in part by NSF under grants CNS-1642982 and CNS-2129015.}
\thanks{The authors are with the School of Electrical, Computer and Energy Engineering, Arizona State University, AZ.}
\thanks{Email: bkeshav1@asu.edu, nicolo.michelusi@asu.edu}

\vspace{-20mm}}
\begin{document}
\maketitle
\thispagestyle{plain}
\pagestyle{plain} 
\setulcolor{red}
\setul{red}{2pt}
\setstcolor{red}

\begin{abstract}
A novel LEarning-based Spectrum Sensing and Access (LESSA) framework is proposed, wherein a cognitive radio (CR)  learns a time-frequency correlation model underlying spectrum occupancy  of licensed users (LUs) in a radio ecosystem; concurrently, it devises \hlt{an approximately optimal spectrum sensing and access policy} under sensing constraints.
A Baum-Welch algorithm is proposed to learn a parametric Markov transition model of LUs' spectrum occupancy based on noisy spectrum measurements. Spectrum sensing and access are cast as a Partially-Observable Markov Decision Process, \hlt{approximately} optimized
 via randomized point-based value iteration.
 \hlt{Fragmentation, Hamming-distance state filters and Monte-Carlo methods are proposed to alleviate the inherent computational complexity,} and
a weighted reward metric to regulate the trade-off between CR's throughput and interference to the LUs. Numerical evaluations demonstrate
that LESSA performs  within 
$5$\% of a genie-aided upper bound with foreknowledge of LUs’ spectrum occupancy, and \hlt{outperforms state-of-the-art algorithms across the entire trade-off region:} $71$\%  over correlation-based clustering, $26$\% over Neyman-Pearson-based spectrum sensing, $6$\%  over the Viterbi algorithm, and \hlt{$9$\% over adaptive Deep Q-Network.} LESSA is then extended to a distributed Multi-Agent setting (MA-LESSA), \hlt{by proposing novel neighbor discovery and channel access rank allocation.} MA-LESSA improves CRs' throughputs by \hlt{$43$\%  over cooperative TD-SARSA}, $84$\% over cooperative greedy distributed learning, and $3\times$ over non-cooperative learning via g-statistics and ACKs. Finally, MA-LESSA is implemented on the DARPA SC2 platform, manifesting superior performance over competitors in a real-world TDWR-UNII WLAN emulation; its implementation feasibility is further validated on an ad-hoc distributed wireless testbed of ESP32 radios, exhibiting $96$\% success probability.
\end{abstract}
\vspace{-5mm}

\section{Introduction}\label{O}
Cognitive radios (CRs) have been touted as instrumental in solving resource-allocation problems in resource-constrained radio environments. Their adaptive computational intelligence facilitates the dynamic allocation of scarce network resources, particularly the spectrum. With the advent of fifth-generation cellular technologies \cite{5GSurvey}, a multitudinous array of devices will be brought into the wireless communication ecosystem, resulting in an enormous strain on the available spectrum resources. Dynamic spectrum access, the key defining feature of CR networks, is being widely studied as a solution to the problem of spectrum scarcity, in both military and consumer spheres: CRs intelligently access portions of the spectrum unused by the sparse and infrequent transmissions of incumbents or licensed users (LUs), in order to deliver their own network flows, while adhering to interference compliance requirements. In order to intelligently access the spectrum white-spaces, CRs need to solve for a spectrum sensing and subsequent access policy based on noisy observations of the occupancy behavior of LUs. Yet, critical design limitations, driven by energy efficiency requirements or constraints on sensing times \cite{WCL:3}, prevent CRs from sensing simultaneously the entire spectrum of interest. Under these constraints, CRs can only sense a small portion of the spectrum to determine access opportunities, as studied in \cite{WCL:4,WCL:10,WCL:11,WCL:3,WCL:5, WCL:6, WCL:8, WCL:9}. However, this approach is quite conservative, since it does not allow CRs to access the large pool of subcarriers that have not been sensed.

LUs' occupancy may exhibit correlation across both time and frequency, as demonstrated in \cite{WCL:12} and visualized in \figref{fig:correlation}. Exploiting this time-frequency correlation structure may significantly improve white-space detection, by enabling CRs to predict the occupancy state of those subcarriers that have not been directly sensed, and may unlock additional spectrum access opportunities. In this paper, focusing first on deployments with a single CR, we propose LESSA,
a LEarning-based Spectrum Sensing and Access framework that leverages these time-frequency correlations: a CR learns a parametric 
time-frequency correlation model underlying the occupancy behavior of LUs in the radio ecosystem; concurrently, it exploits this learned model to construct \hlt{an approximately optimal sensing and access policy using a Partially Observable Markov Decision Process (POMDP) formulation, solved via randomized Point-Based Value Iteration (PBVI). 
We propose fragmentation techniques, Hamming distance state filters 
and Monte-Carlo methods to alleviate the inherent computational complexity,
and we introduce a weighted reward metric to regulate the trade-off between CR's throughput and interference to the LUs, unlike the state-of-the-art.
 Our numerical evaluations demonstrate that LESSA improves spectrum occupancy across the entire trade-off region, over state-of-the-art algorithms including
correlation-based clustering \cite{WCL:7}, 
 Neyman-Pearson-based spectrum sensing \cite{WCL:11}, 
 the Viterbi algorithm \cite{WCL:6}, and adaptive  Deep Q-Network (DQN) \cite{WCL:DQN}.
Furthermore, we extend LESSA to Multi-Agent (MA-LESSA) deployments, enveloping radio environments with several CRs performing intelligent spectrum sensing and access.
We propose novel neighbor discovery and channel access rank allocation schemes that enable the CRs to collaborate and capitalize on spectral resources left unused by the multiple LUs in the network. 
We implement MA-LESSA  on  the  DARPA Spectrum  Collaboration  Challenge (SC2)  platform,  manifesting  superior  performance  over  competitors  in  a real-world  TDWR-UNII  WLAN  emulation \cite{DARPA:ActiveIncumbent}.
Finally, we demonstrate the implementation feasibility of MA-LESSA in a network of ESP32 radios, and we illustrate the performance disparities between collaborative and opportunistic (non-cooperative or competitive) access through comparisons with algorithms in the multi-agent state-of-the-art.}

{\bf Related Work:} \hlt{We now discuss the most closely-related works in the state-of-the-art, summarized in Table~\ref{tab:comp}.}
Most prior  algorithms for  spectrum sensing and access operate under the assumption that the occupancy behavior of LUs is
independent across both time and frequency \cite{WCL:4, WCL:10, WCL:3, WCL:11, WCL:MIT},
or exploit temporal correlation but fail to 
capitalize on frequency correlation \cite{WCL:5, WCL:6, WCL:8, WCL:9}.
These assumptions are not only impractical \cite{WCL:12} but also imprudent because critical information aiding the accurate detection of white-spaces may be gleaned by exploiting the correlation in LUs' occupancy over \emph{both} time and frequency, as done in this work. 

Spectrum sensing algorithms exploiting both time-frequency correlation in LU occupancy are studied in \cite{WCL:7, WCL:DQN, LSTM, CNN}. Yet, the system detailed in \cite{WCL:7} learns the time-frequency LU occupancy correlation structure 
 offline using pre-loaded databases, which may be inefficient in non-stationary settings; in contrast,
 we propose a \emph{concurrent learning and adaptation framework}, in which
CRs learn a  time-frequency correlation model of LUs' occupancy via an online Baum-Welch algorithm, and leverage this knowledge to concurrently optimize spectrum sensing and access (under sensing limitations). \hlt{On the other hand, LESSA achieves performance superior to model-free DQN \cite{WCL:DQN}, owing to a parametric time-frequency Markov transition model. Contrasting our framework against black-box Deep Learning models \cite{LSTM, CNN}, our approach circumvents laborious data collection and pre-processing tasks, thanks to online learning of the parametric time-frequency correlation model underlying the LUs' occupancy behavior.}

In terms of the observation model, a noiseless spectrum sensing setting is assumed in \cite{WCL:7}, which is unrealistic. In \cite{WCL:5}, LUs' spectrum occupancy is estimated directly via energy detection thereby ignoring errors in state estimation, while we incorporate a multi-subcarrier access decision based on the contemporary posterior belief probability distribution. \hlt{Although the DQN framework in \cite{WCL:DQN} revolves around a POMDP formulation, the uncertainty involved is assumed to be only due to sensing restrictions by the CR, while making no claims about system operations in noisy settings.} Differently from these works, we assume a Hidden Markov Model (HMM) formulation in which the true LU occupancy states are hidden behind noisy observations at a CR's spectrum sensor.
Hence, partial observability in our formulation is due simultaneously to channel sensing restrictions, a noisy observation model, and unknown LU occupancy dynamics.
\begin{table}
\begin{center}
\hlt{\scriptsize
\begin{tabular}{| c | c | c | c | c | c | c | c |}
\hline
\thead{\bf Paper}&
\thead{\bf \!\!\!Adaptive\!\!\!\vspace{-1.3mm}\\sensing} & 
\thead{\bf Sensing\vspace{-1.3mm}\\\!\!\!constraints\!\!\!} & 
\thead{\bf \!\!\!Training mode:\!\!\!\vspace{-1.3mm}\\Model/Policy} & 
\thead{\bf LUs' occupancy\vspace{-1.3mm}\\\!\!\!correlation model\!\!\!} & 
\thead{\bf \!\!\!Observation\!\!\!\vspace{-1.3mm}\\model} & 
\thead{\bf \!\!\!CR-LU trade-off\!\!\!\vspace{-1.3mm}\\regulation} & 
\thead{\bf \!\!\!CRs' deployment\!\!\!}
\\\hline
  \!\!\!This work\!\!\! & Yes & Yes & Online/Online & Time-Freq., parametric & Noisy & Yes & Single- \& Multi-
    \\\hline
 \cite{WCL:DQN} & Yes & Yes & Online/Online & Time-Freq., Model-free & Noise  ignored & No & Single-
   \\\hline
  \cite{WCL:7} & Yes & Yes & Offline/Offline & Time-Freq., parametric & Noiseless & No & Single-
  \\\hline
  \cite{LSTM} & No & No & Offline/Offline & Time-Freq., Model-free & Noiseless & No & Single-
  \\\hline
  \cite{CNN} & No & Yes & Offline/Offline & Time-Freq., Model-free & Noisy & No & Multi-
  \\\hline
  \cite{WCL:5} & Yes & Yes & Online/Online & Time only & Noisy & No & Multi-
  \\\hline
  \cite{WCL:8} & Yes & Yes & Known/Online & Time only & Noisy & No & Single-
  \\\hline
  \cite{WCL:9} & Yes & Yes & Online/Online & Time only & Noisy & No & Multi-
  \\\hline
  \cite{WCL:6} & No & No & Offline/Offline & Time only & Noisy & No & Single-
  \\\hline
  \cite{WCL:4} & Yes & Yes & -/Online & Independence & Noisy & No & Single-
  \\\hline
    \cite{WCL:MIT} & Yes & Yes & -/Online & Independence & Noisy & No & Multi-
  \\\hline
  \cite{WCL:10} & No & Yes & -/Online & Independence & Noisy & No & Multi-
  \\\hline
   \cite{WCL:11} & No & No & -/- & Independence & Noisy & No & Multi-
  \\\hline
  \cite{WCL:3} & No & Yes & -/- & Independence & Noisy & No & Multi-
  \\\hline
\end{tabular}
}
\vspace{-2mm}
\caption{Comparison of our framework with relevant schemes in the state-of-the-art.\label{tab:comp}}
\vspace{-14mm}
\end{center}
\end{table}

\vspace{-7mm}
All the above works fail to provide a mechanism to manage the trade-off between CRs' network throughput and LUs' interference; in contrast, LESSA enables this feature through a weighted reward metric that favors CRs' throughput and penalizes LUs' interference. Unlike non-adaptive sensing strategies \cite{WCL:3, WCL:6, WCL:10, WCL:11, LSTM, CNN}, our solution adapts the sensing action in each time-step to more effectively cope with spectrum sensing constraints, driven by transition model estimates and reward/penalty feedback. 

Finally, analyzing the state-of-the-art in the multi-agent distributed CR domain, we find both collaborative \cite{WCL:3, WCL:5, WCL:MIT, WCL:9, WCL:10, WCL:11, CNN} as well as opportunistic \cite{WCL:MIT} schemes for channel sensing and access. Drawing a contrast between our contributions and the systems detailed in these works, the authors in \cite{WCL:9} make independence assumptions during a projection approximation to a lower-dimensional space; \cite{WCL:3, WCL:11} assume a time-frequency independence structure in LUs' occupancy, and employ energy detection at the CRs, i.e., no adaptive sensing and more importantly, no policy optimization; 
\cite{WCL:10} focuses primarily on data aggregation strategies within the ensemble; 
\cite{WCL:5}
proposes a  multi-agent Temporal Difference (TD) SARSA with Linear Function Approximation (LFA), but
fails to detail neighbor discovery and channel access order allocation schemes, proposed in this paper; \cite{WCL:MIT} details a collaborative scheme (greedy learning under pre-allocation) and an opportunistic one (g-statistics with ACKs), under the assumption of
 time-frequency independence in LUs'  occupancy behavior; additionally, the opportunistic scheme in \cite{WCL:MIT} relies on ACKs as a feedback mechanism to gauge the utility of an access decision, which imbibes unnecessary lag into the model; in contrast, our framework employs a threshold-based decision heuristic involving the posterior belief probability to evaluate the reward obtained from the executed access action: in addition to displaying superior performance, as illustrated in \secref{III}, this mechanism is easier to implement in real-world settings, as we demonstrate by realizing our solution on the DARPA SC2 emulator \cite{DARPA:SC2} and on an ad-hoc ESP$32$ network \cite{Espressif:ESP32}.

{\bf Contributions:} In a nutshell, the contributions of this paper are summarized as follows:
\label{itemcontribs}
\begin{itemize}[leftmargin=*]
    \item We develop a LEarning-based framework for Spectrum Sensing and Access (LESSA) in a radio environment with LUs exhibiting time- and frequency- Markovian correlation in their occupancy behavior, and a CR attempting to  detect and access unused spectral resources -- under a noisy observation model with sensing restrictions.
    \item In \secref{II.I}, we develop an online
    Baum-Welch algorithm \cite{Rabiner_1989}
     to learn the LUs' occupancy correlation model.
    \item \hlt{In \secref{II.II}, we concurrently leverage the learned model in a randomized PBVI algorithm known as PERSEUS \cite{WCL:13} to devise an approximately optimal spectrum sensing and access policy; we alleviate its computational complexity by introducing fragmentation, belief update simplification heuristics via Hamming distance state filters and Monte-Carlo based methods.}
    \item In single-agent settings, we demonstrate the superior performance of LESSA relative to relevant algorithms in the state-of-the-art (\secref{III}). \hlt{Additionally, through computational time complexity bench-marking, we prove the enhanced scalability of our solution.}
    \item In \secref{Z}, we extend LESSA to distributed Multi-Agent (MA-LESSA) deployments: \label{mdware}\hlt{adapting the Multi-band Directional Neighbor Discovery scheme described in \cite{MDND} to distributed multi-agent CR deployments, we propose a novel neighbor discovery heuristic centered around RSSI thresholding; inspired by cluster fallback mechanisms (leader selection, broker failover, master-slave auto-configuration, and data replication) involved in Message Oriented Middleware \cite{AMQ}, we propose a channel access rank allocation technique centered around a quorum-based preferential ballot algorithm.} On this front, we demonstrate improved performance over both collaborative and opportunistic distributed multi-agent state-of-the-art; also, we exemplify its implementation feasibility on an ad-hoc WLAN testbed of ESP32 radios \cite{GCTronic:epuck2, Espressif:ESP32}.
    \item Finally, for multi-agent evaluations, we retrofit our proposed solution into the DARPA SC2 BAM! Wireless radio \cite{BAM} to emulate its operations during the Active Incumbent scenario (TDWR-UNII WLAN) \cite{DARPA:ActiveIncumbent}, and prove superior scoring performance over competing strategies 
     \cite{DARPA:CIL, DARPASC2:end1, 8935729, DARPASC2:end3,8935774}. \hlt{We also perform computational time complexity analyses of our neighbor discovery and channel access rank allocation heuristics in emulations of highly-mobile real-world disaster relief (SC$2$ Payline \cite{DARPA:Payline}) and military deployment scenarios (SC$2$ Alleys of Austin \cite{DARPA:Alleys}).}
\end{itemize}

The rest of this paper is organized as follows: \secref{t} details the system model; \secref{II} describes our algorithmic solutions; \secref{III} presents numerical evaluations for the single-agent case; \secref{Z} extends LESSA to a distributed multi-agent setup, with numerical evaluations; finally, \secref{V} provides concluding remarks.
\vspace{-3mm}
\section{System Model}\label{t}
\vspace{-2mm}
\subsection{Signal Model}\label{I.I}
\begin{figure} [t]
    \centerline{
    \includegraphics[width=0.7\linewidth]{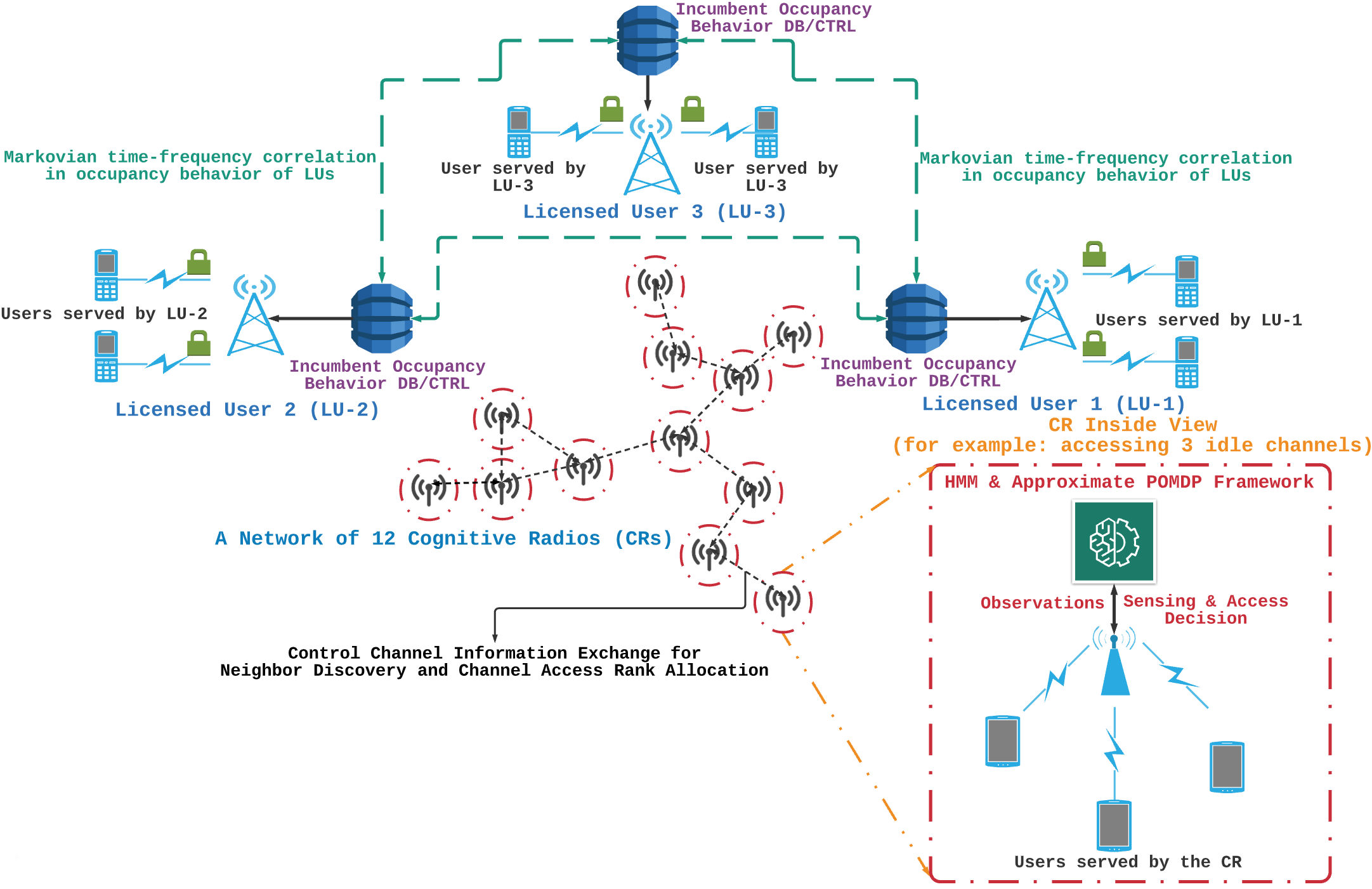}}
    \vspace{-4mm}
    \caption{The radio ecosystem under analysis: An exemplification of the system model detailed in \secref{I.I} with $J_L{=}3$ and $J_C{=}12$: we first study deployments with $J_C{=}1$ before extending our analyses to multi-agent settings.}
    \label{fig: A.0}
    \vspace{-8mm}
\end{figure}

We consider a licensed network of $J_L$ LUs, and $J_C$ CRs attempting to exploit portions of the spectrum left unused by these LUs -- as illustrated in \figref{fig: A.0}. In Sections~\ref{t}, \ref{II}, and \ref{III}, we focus on the single-agent case ($J_C{=}1$); the multi-agent extension ($J_C{>}1$) is discussed in \secref{Z}. The spectrum of interest is discretized into $K$ subcarriers of bandwidth $W$. The frequency-domain signal received at the CR's spectrum sensor in time-slot $t$ at subcarrier $k$ is 
\begin{align}\label{1}
    Y_{k}(t)=\sum_{j{=}1}^{J_L}{H_{j,k}(t)X_{j,k}(t)+V_{k}(t)},
\end{align}
where $X_{j,k}(t)$ is the frequency-domain signal of LU $j{\in}\{1,{\dots},J_L\}$ on subcarrier $k \in \{1,{\dots},K\}$, with $X_{j,k}(t){=}0$ if LU $j$ is idle on subcarrier $k$; $H_{j,k}(t)$ denotes the channel on subcarrier $k$ between LU $j$ and the CR; and $V_{k}(t){\sim}\mathcal{CN}(0,\sigma_{V}^{2})$
denotes Gaussian noise with variance $\sigma_{V}^{2}$, i.i.d across time and subcarriers. We assume an Orthogonal Frequency Division Multiple Access (OFDMA) strategy among the LUs:
letting $j_{k,t}$ be the index of the LU that occupies subcarrier $k$ in time-slot $t$,  $X_{k}(t){\triangleq}X_{j_{k,t},k}(t)$ and $H_{k}(t){\triangleq}H_{j_{k,t},k}(t)$, we can rewrite \eqref{1} as
\begin{align}\label{2}
    Y_{k}(t)=H_{k}(t)X_{k}(t)+V_{k}(t),
\end{align}
where $X_{k}(t){=}0$ if subcarrier $k$ is not occupied in time-slot $t$. We model $H_{k}(t)$ as Rayleigh fading with variance $\sigma_{H}^{2}$: $H_{k}(t){\sim}\mathcal{CN}(0,\sigma_{H}^{2})$, i.i.d across time and subcarriers.
\vspace{-4mm}
\subsection{Time-Frequency Occupancy Correlation Structure}\label{I.II}
\begin{figure} [t]
     \begin{subfigure}{0.43\linewidth}
         \centering
         \includegraphics[width=0.9\linewidth]{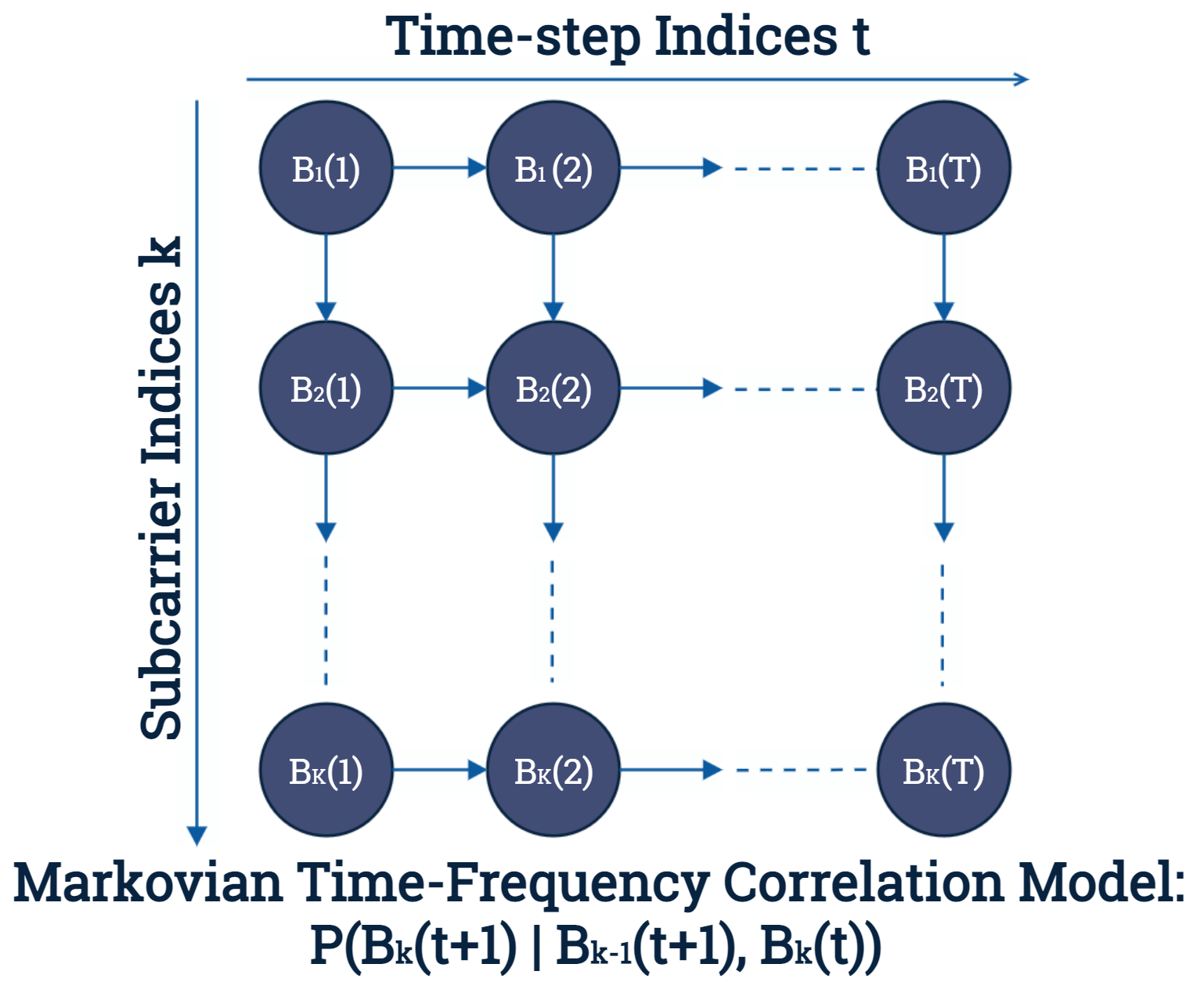}
         \caption{Markovian time-frequency correlation model}
         \label{fig:correlation}
     \end{subfigure}
     \begin{subfigure}{0.57\linewidth}
         \centering
         \includegraphics[width=0.9\linewidth,trim=0 0 0 10,clip]{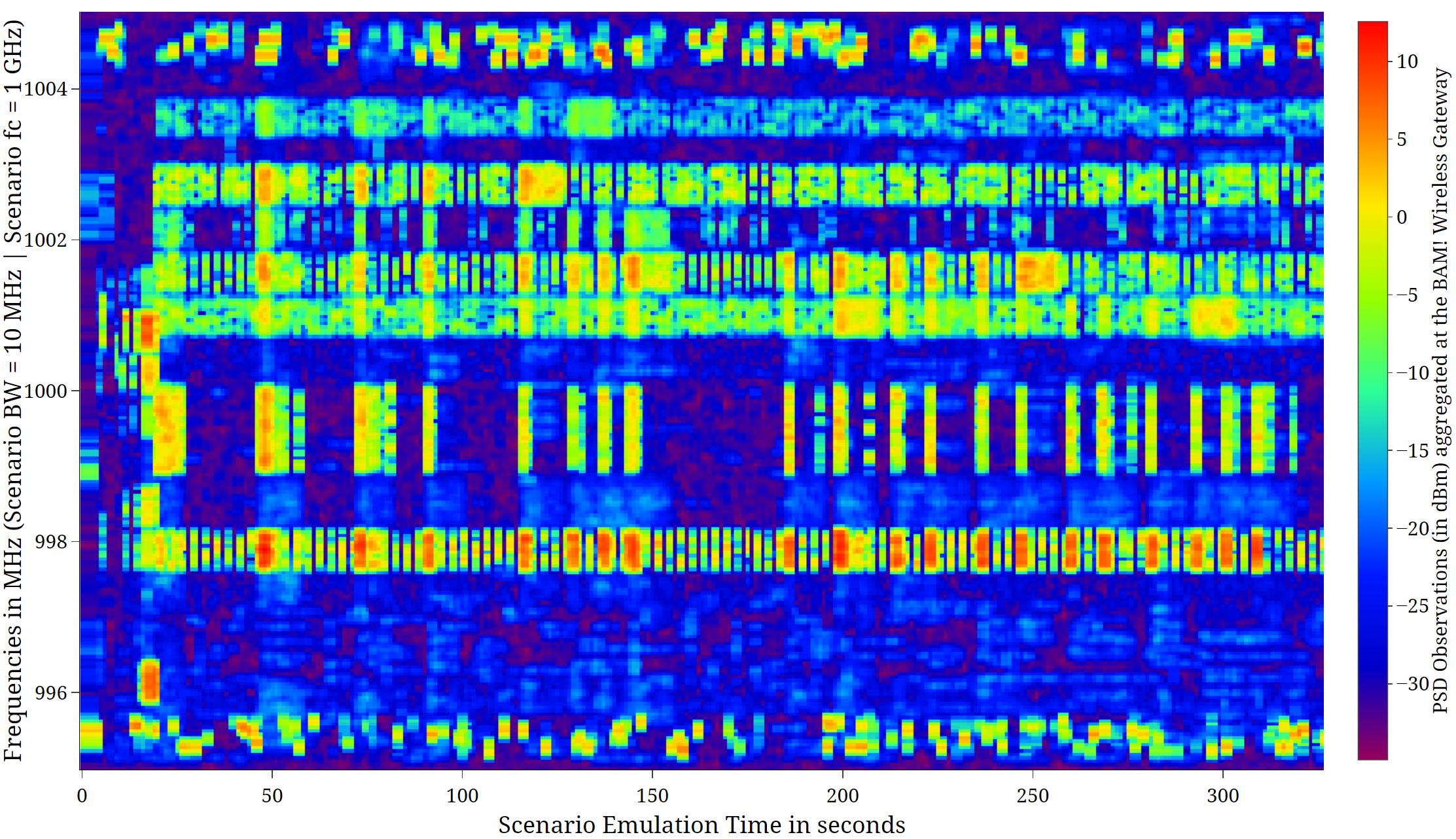}
         \caption{Active Incumbent PSD data}
         \label{fig:psd}
     \end{subfigure}
     \vspace{-4mm}
     \caption{Visualization of the LU occupancy time-frequency correlation structure (a); combined PSD plot of the occupancy behavior of a TDWR and competitors during the DARPA SC2 Active Incumbent scenario emulation (b).}
     \label{fig:corr_and_psd}
     \vspace{-8mm}
\end{figure}

The LU's signal on subcarrier $k$ in time-slot $t$ is modeled as
\begin{align}\label{3}
    X_{k}(t)=\sqrt{P_{T}}B_{k}(t)S_{k}(t),
\end{align}
where $P_{T}$ denotes the transmission power of the occupant LU; $B_{k}(t)$ represents the binary occupancy variable, with $B_{k}(t){=}1$ if subcarrier $k$ is occupied by a LU in time-slot $t$, and $B_{k}(t){=}0$ otherwise; $S_{k}(t)$ is the transmitted symbol, i.i.d across time and subcarriers, modeled from a certain constellation with $\mathbb{E}[|S_{k}|^{2}]=1$. Then, $H_{k}(t)X_{k}(t){=}\sqrt{P_{T}}B_{k}(t)H_{k}(t)S_{k}(t)$. Herein, we approximate $H_{k}(t)S_{k}(t)$ as a zero-mean complex Gaussian random variable with variance $\sigma_{H}^{2}$. We denote the spectrum occupancy state in time-slot $t$ as
\begin{align}\label{4}
    \vec{B}(t)=[B_{1}(t),B_{2}(t),B_{3}(t),\dots,B_{K}(t)]^{\intercal} \in \{0,1\}^{K}.
\end{align}
We assume that spectrum occupancy is correlated across time and subcarriers: LUs typically occupy a set of adjacent subcarriers (frequency correlation), repeating similar motifs in behavior over an extended period of time (temporal correlation) \cite{WCL:12, 4213046,McHenry:2006:CSO:1234388.1234389}. \hlt{To capture temporal correlation, we model the evolution of $\vec{B}(t)$ over time as a Markov process
\begin{align}\label{5}
    \mathbb{P}(\vec{B}(t+1)|\vec{B}(j),\forall j \leq i)=\mathbb{P}(\vec{B}(t+1)|\vec{B}(t))\triangleq P_B(\vec{B}(t+1)|\vec{B}(t)),
\end{align}
\label{page:corrmodel}
with one-step transition probability $P_B(\vec{B}(t+1)|\vec{B}(t))$.
Using the chain rule of conditional probability, we can further express it as
\begin{align}
\label{condprob}
P_B(\vec{B}(t{+}1)|\vec{B}(t))
=
\mathbb{P}(B_1(t{+}1)|\vec{B}(t))
\prod_{k=2}^K\mathbb{P}(B_k(t{+}1)|\vec{B}_{1:k-1}(t{+}1),\vec{B}(t)),
\end{align}
where $\vec{B}_{1:k-1}(t+1)$ is the spectrum occupancy state in subcarriers $1$ to $k-1$ in time-slot $t+1$.
We further assume a Markovian structure across frequency,
\begin{align*}
    \mathbb{P}(B_{k}(t{+}1){|}\vec{B}_{1:k-1}(t{+}1),\vec{B}(t))=\mathbb{P}(B_{k}(t{+}1){|}B_{k-1}(t{+}1),\vec{B}(t)),
\end{align*}
i.e., 
$B_{k}(t{+}1)$ is independent of the spectrum occupancy in the non-adjacent subcarriers $\vec{B}_{1:k-2}(t{+}1)$,
when conditioned on $\vec{B}(t)$ and on the
    spectrum occupancy in the adjacent subcarrier $B_{k-1}(t{+}1)$.
 This assumption reflects the intuition that 
    the state of the adjacent subcarrier $k-1$ ($B_{k-1}(t{+}1)$) more directly affects the state of subcarrier $k$ than non-adjacent subcarriers $1$ to $k-2$.
    Replacing this expression into \eqref{condprob}, we finally obtain
    \begin{align}
    \label{6}
P_B(\vec{B}(t{+}1)|\vec{B}(t))
=
\mathbb{P}(B_1(t{+}1)|\vec{B}(t))
\prod_{k=2}^K\mathbb{P}(B_k(t{+}1)|B_{k-1}(t{+}1),\vec{B}(t)).
\end{align}
We denote this model as \emph{bottom-up frequency correlation}, since 
the state of subcarrier $k$ depends on that of the adjacent lower subcarrier $k-1$,
as opposed to the \emph{top-down frequency correlation} where 
 it depends on that of the adjacent upper subcarrier $k+1$.
We remark that bottom-up and top-down frequency correlation models can be used interchangeably: in fact,
replacing
$\mathbb{P}(B_k(t{+}1)|B_{k-1}(t{+}1),\vec{B}(t))=
\mathbb{P}(B_{k-1}(t{+}1)|B_{k}(t{+}1),\vec{B}(t))\frac{\mathbb{P}(B_{k}(t{+}1)|\vec{B}(t))}{\mathbb{P}(B_{k-1}(t{+}1)|\vec{B}(t))}$ (Bayes' rule)
in \eqref{6} and simplifying, we obtain the top-down frequency correlation model
\begin{align}
P_B(\vec{B}(t{+}1)|\vec{B}(t))
=
\mathbb{P}(B_{K}(t{+}1)|\vec{B}(t))
\prod_{k=1}^{K-1}
\mathbb{P}(B_{k}(t{+}1)|B_{k+1}(t{+}1),\vec{B}(t)),
\end{align}
so that the two models can be directly mapped to each other.
We now embed the bottom-up frequency correlation of \eqref{6} into a parametric form.
Expressing $\vec{B}(t)=[B_k(t),\vec{B}_{-k}(t)]$, where 
$\vec{B}_{-k}(t)$ is the spectrum occupancy state in subcarriers other than $k$ in time-slot $t$,
we define
    \begin{align}
    \label{7}
    q_{w}&\triangleq\mathbb{P}(B_{1}(t{+}1)=1|B_{1}(t){=}w,\vec{B}_{-1}(t)):\forall w \in \{0,1\},\\
        p_{u,v}&\triangleq\mathbb{P}(B_{k}(t{+}1){=}1|B_{k-1}(t{+}1){=}u,B_{k}(t){=}v,\vec{B}_{-k}(t)):\forall u,v \in \{0,1\}, 2{\leq}k{\leq}K.
 \noindent
    \end{align}
Note that this parametric model assumes that $B_k(t{+}1)$ is independent of $\vec{B}_{-k}(t)$, when conditioned on 
$(B_{k-1}(t{+}1),B_k(t))$: intuitively, the state of subcarrier $k$ in time-slot $t{+}1$ is most directly affected by the state on the same subcarrier in the previous time-slot $t$, as shown in \figref{fig:correlation}.
In the following,
we define the parameter vector $\vec{\theta}=[p_{0,0},p_{0,1},p_{1,0},p_{1,1},q_{0},q_{1}]$,
and express 
the dependence of the one-step transition model on $\vec\theta$ as
$P_B(\vec{B}(t{+}1)|\vec{B}(t);\vec\theta)$.}

To experimentally validate the aforementioned parametric time-frequency correlation model, we performed a Bayesian Information Criterion (BIC) evaluation \cite{BIC} on a dataset of Power Spectral Density (PSD) measurements
 from the DARPA SC2 Active Incumbent scenario emulation \cite{DARPA:ActiveIncumbent}, depicted in \figref{fig:psd}. This scenario constitutes
 a Terminal Doppler Weather Radar (TDWR) station, and several competitor CR networks \cite{DARPASC2:end3,8935774} (including our Purdue BAM! Wireless \cite{BAM}) of Unlicensed National Information Infrastructure (UNII: $5$GHz WLAN) Wi-Fi nodes.
 We define the BIC metric as
 $\mathrm{BIC} = \Gamma\cdot\ln{\nu} - 2\ln{\mathbb{P}(\mathbf{B}|\vec{\theta}^*)}$,
where $\nu$ is the sample size, \hlt{$\Gamma$ is the number of model parameters} (6 in the proposed parametric model), $\mathbf{B}$ is the time-frequency binary occupancy matrix from the dataset, and $\vec{\theta}^*$ is the  
model parameter vector estimated from the dataset of PSD measurements, using the Baum-Welch algorithm detailed in \secref{II.I}. We employed a $70{-}30$ training-test split to evaluate this metric, i.e., the occupancy data collected during the first $70$\% of the $330$ seconds of scenario emulation is used to estimate the model parameters, while the remaining $30$\% is dedicated to the BIC evaluation. The parametric bottom-up time-frequency correlation model yields a BIC of $71.872$, the best compared to other state-of-the-art models: time-frequency independence ($98.840$) \cite{WCL:3, WCL:4, WCL:MIT, WCL:10, WCL:11}; time-only correlation ($95.231$) \cite{WCL:5, WCL:6, WCL:8, WCL:9}; frequency-only correlation  ($76.879$);
\hlt{notably, the \emph{top-down} frequency correlation model, using a similar parameterization of \eqref{7}, yields a similar BIC metric of $74.207$ (the slightly different value is due to the parametric structure).}
This evaluation reveals that exploiting frequency correlation is more important than time correlation -- and not surprisingly, exploiting \emph{both} provides a better fit to the dataset. This assessment validates our conjecture that LUs in real-world  deployments exhibit prominent spectrum occupancy patterns across both time and frequency, exploited by LESSA. 
\hlt{Moreover, it further corroborates our previous observation that
bottom-up and top-down correlation models can be used interchangeably.}
\vspace{-6mm}
\subsection{Channel Sensing Model}\label{I.III}
The CR attempts to detect white-spaces and access them to deliver its network flows. Due to constraints on energy-efficiency and sensing/data aggregation times \cite{WCL:3}, it can sense a maximum of $\kappa$ subcarriers in a time-slot, with $1{\leq}\kappa{\leq}K$. Let $\mathcal{K}_{t}{\subseteq}\{1,2,{\dots},K\}$ be the set of subcarriers sensed by the CR at time $t$, with $|\mathcal{K}_{t}|{\leq}\kappa$. This selection is dictated by a sensing policy, described in \secref{II.0}. After sensing the subcarriers listed in $\mathcal{K}_{t}$, the observation vector  $\vec{Y}(t){=}[Y_{k}(t)]_{k{\in}\mathcal{K}_{t}}$ is collected, with $Y_{k}(t)$ given in \eqref{2}. Owing to \eqref{2}, the  i.i.d. assumptions on the noise $V_{k}(t)$, the transmitted symbols $S_{k}(t)$, and the frequency domain subcarriers $H_{k}(t)$, the probability density function (pdf) of $\vec{Y}(t)$ conditioned on $\vec{B}(t)$ and $\mathcal{K}_{t}$ is given by
\begin{align}\label{8}
    f(\vec{Y}(t)|\vec{B}(t),\mathcal{K}_{t})=\prod_{k\in\mathcal K_t}f(Y_{k}(t)|B_{k}(t)),\text{ where } Y_{k}(t)|B_{k}(t)\sim\mathcal{CN}(0,\sigma_{H}^{2}P_{T}B_{k}(t)+\sigma_{V}^{2}).
\end{align}
\subsection{POMDP Formulation and Spectrum Access}\label{II.0}
\begin{figure} [t]
    \centerline{
    \includegraphics[width=0.7\linewidth]{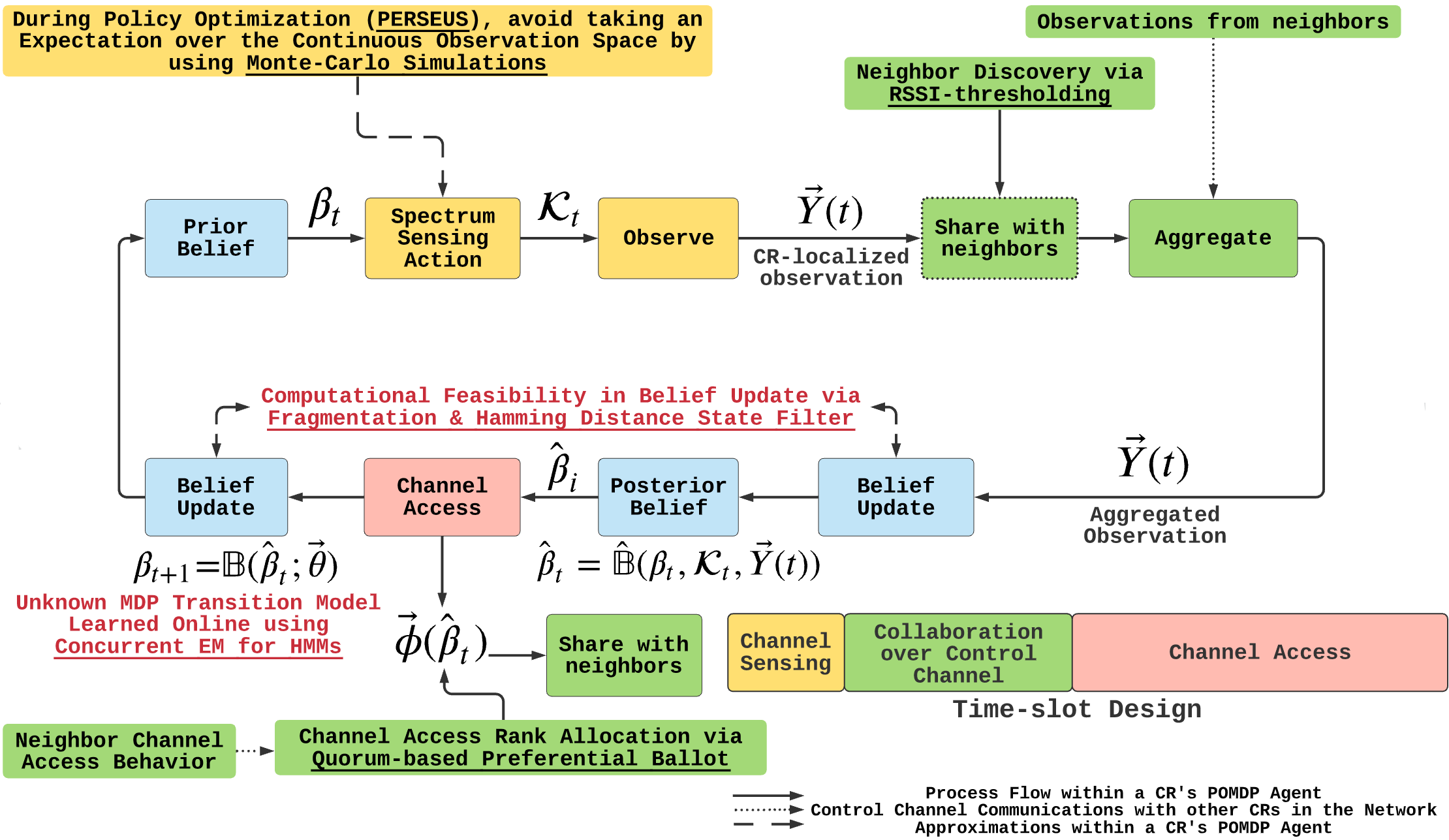}}
    \vspace{-2mm}
    \caption{The POMDP process flow as discussed in \secref{II.0} -- with neighbor discovery, channel access rank allocation, and time-slot design (shown in green) discussed in our multi-agent deployment analysis (\secref{Z}).}
    \vspace{-10mm}
    \label{fig: A.add-1}
\end{figure}

POMDPs model the repeated, sequential interactions of an agent tasked with maximizing its reward, with a stochastic environment, in which the agent has only access to noisy observations. We now describe the POMDP operation, whose process flow is illustrated in \figref{fig: A.add-1} (the multi-agent features, shown in green, will be discussed in \secref{Z}).

Prior to gathering spectrum measurements in time-slot $t$, the POMDP state is given by the prior belief $\beta_{t}$, i.e., the probability distribution of the unknown spectrum occupancy state $\vec{B}(t)$, given the history of measurements obtained by the CR's spectrum sensor up to, but not including, time-slot $t$.
 Given $\beta_{t}$, the CR chooses a sensing action according to a sensing policy $\pi$, $\mathcal{K}_{t}=\pi(\beta_{t}){\in}\mathcal{A}$, 
 where $\mathcal{A}$ is the spectrum sensing space, i.e., all possible combinations in which $\kappa$ subcarriers are chosen to be sensed in a time-slot (discussed in \secref{I.III});
 the CR then
senses the subcarriers in $\mathcal{K}_{t}$, observes $\vec{Y}(t)$, and computes the posterior belief of $\vec{B}(t)$ using Bayes' rule, as
\begin{align}\label{10}
    \begin{aligned}
        \hat{\beta}_{t}(\vec{B})&=\mathbb{P}(\vec{B}(t)=\vec{B}|\beta_{t},\mathcal{K}_{t},\vec{Y}(t))
        =\frac{f(\vec{Y}(t)|\vec{B},\mathcal{K}_{t})\beta_t(\vec{B})}{\sum_{\vec{B}' \in \{0,1\}^{K}}f(\vec{Y}(t)|\vec{B}',\mathcal{K}_{t})\beta_{t}(\vec{B}')}.
    \end{aligned}
\end{align}
Given  $ \hat{\beta}_{t}$, the CR then makes channel access decisions
$\vec{\phi}{\in}\{0,1\}^K$, where
$\phi_k{=}1$ denotes the CR's decision to access the $k$th subcarrier, otherwise $\phi_k{=}0$.
We design $\vec{\phi}$ based on the reward metric
\begin{align}\label{12a}
    R(\vec{\phi},\vec{B}(t))=\sum_{k=1}^{K}(1-B_{k}(t))\phi_{k}-\lambda B_{k}(t)\phi_k.
\end{align}
Since $\vec{B}(t)$ is unknown, we compute its
 expectation based on the posterior belief $\hat{\beta}_{t}$ as
\begin{align}\label{12}
    R(\vec{\phi},\hat{\beta}_{t})=\mathbb{E}[R(\vec{\phi},\vec{B}(t))|\vec{\phi},\hat{\beta}_{t}]=\sum_{k=1}^{K}(1-\hat{\beta}_{t,k})\phi_{k}-\lambda \hat{\beta}_{t,k}\phi_k,
\end{align}
where $\hat{\beta}_{t,k}=\sum_{\vec{B}\in\{0,1\}^K: B_k=1}\hat{\beta}_{t}(\vec{B})$
is the posterior probability that subcarrier $k$ is occupied by a LU.
Note that, if the CR uses the $k$th subcarrier ($\phi_{k}(t)=1$), it accrues a reward of $1$ if the subcarrier is truly idle (with posterior probability $1-\hat{\beta}_{t,k}$), and a penalty of $\lambda\geq 0$ if it is occupied (with posterior probability $\hat{\beta}_{t,k}$); it accrues no reward for not using a channel. Hence, \eqref{12}
balances a trade-off between maximizing the CR's throughput
(CR's transmissions on idle subcarriers), and minimizing the interference to the LUs (CR's transmissions on occupied ones); $\lambda$ regulates such trade-off. The channel access decision is obtained as $\vec{\phi}^{*}(\hat{\beta}_{t}){=}\arg\max_{\vec{\phi}\in\{0,1\}^K} R(\vec{\phi},\hat{\beta}_{t})$, yielding
\begin{align}
\label{SAdecision}
\phi_k^{*}(\hat{\beta}_{t})
=\mathcal I[\hat{\beta}_{t,k}{\leq}(1{+}\lambda)^{-1}],\ \forall k=1,\dots,K,
\end{align}
where $\mathcal I[\cdot]$ is the indicator function,
yielding the optimized reward
\begin{align}
\label{R*}
R^*(\hat{\beta}_{t})=\max_{\vec{\phi}\in\{0,1\}^K} R(\vec{\phi},\hat{\beta}_{t})
=
\sum_{k=1}^{K}\max\{1-(1+\lambda)\hat{\beta}_{t,k},0\}.
\end{align}
In other words, if the CR is confident that the $k$th subcarrier is idle ($\hat{\beta}_{t,k}{<}\frac{1}{1{+}\lambda}$), it accesses it; otherwise, it remains idle. The confidence level $\frac{1}{1{+}\lambda}$ is regulated by the penalty parameter $\lambda$.

After accessing the subcarriers based on the access decision $\phi_{k}^{*}(\hat\beta_t)$, the CR then
updates the prior belief for the next time-slot $t+1$ based on the one-step transition model,
\begin{align}\label{13}
    \beta_{t+1}(\vec{B})=\sum_{\vec{B}'\in\{0,1\}^K}P_B(\vec{B}|\vec{B}';\vec\theta)\hat{\beta}_t(\vec{B}'),
\end{align}
and the process is repeated over time.
Let
\begin{align}\label{14}
    \hat{\beta}_{t}=\hat{\mathbb{B}}(\beta_{t},\mathcal{K}_{t},\vec{Y}(t)),\ \ \ \beta_{t+1}=\mathbb{B}(\hat{\beta}_{t};\vec\theta)
\end{align}
denote the functions that map the prior belief $\beta_{t}$ to the posterior belief $\hat{\beta}_{t}$, and the latter to the next prior belief $\beta_{t+1}$ (see Eqs. \eqref{10} and \eqref{13}). The goal is to devise a spectrum sensing policy (the spectrum access policy is given in \eqref{SAdecision}) to maximize the infinite-horizon discounted reward
\begin{align}\label{17}
    V^{\pi}(\beta)=\mathbb{E}_{\pi}\left[\sum_{t=1}^{\infty}\gamma^{t}
    R^*(\hat{\beta}_{t})\Big{|}\beta_{0}=\beta\right],
\end{align}
where
 $0{<}\gamma{<}1$ is the discount factor, $\beta_{0}{=}\beta$ (e.g., uniform) is the initial belief, and $\hat{\beta}_{t}$ is the posterior belief induced by the policy $\mathcal{K}_{t}{=}\pi(\beta_{t})$ and the observation vector $\vec{Y}(t)$ via $\hat{\beta}_{t} = \hat{\mathbb{B}}(\beta_{t}, \mathcal{K}_{t},\vec{Y}(t))$.
 The maximization problem is then stated as
 \begin{align}\label{16}
    \pi^{*}=\argmax_{\pi}V^{\pi}(\beta),
\end{align}
yielding the optimal value $V^*(\beta)$ under the optimal spectrum sensing policy $\pi^*$.
  In principle, $V^*(\beta)$ can be determined via the value iteration algorithm
  $V_{i+1}=\mathcal H(V_i)$, 
  initialized as $V_0(\beta)=0$,
  which converges to  $V^*$ as $i\to\infty$ \cite{PUOccupancy:18}.
  Here, $\mathcal H$ is the Bellman's operator, defined as
    \begin{align}
    \label{18}
        V_{i+1}(\beta)
        =\max_{\mathcal{K} \in \mathcal{A}}\mathbb{E}_{\vec{Y}|\vec{B}\sim\beta,\mathcal{K}}\Big[R^*(\overbrace{\hat{\mathbb{B}}(\beta,\mathcal{K},\vec{Y})}^{\hat\beta})+\gamma\cdot V_{i}\Big(\overbrace{\mathbb{B}(\hat{\mathbb{B}}(\beta,\mathcal{K},\vec{Y});\vec\theta)}^{\beta_{next}}\Big)\Big],\ \forall \beta.
    \end{align}
    \hlt{We term \eqref{18} the \emph{backup} operation, which in exact value iteration should be solved for every belief $\beta$ in the probability probability simplex.
    Intuitively, $V_{i}(\beta)$ and $V_{i+1}(\beta)$ represent the optimal finite horizon (of duration $i$ and $i+1$, respectively) discounted values starting from belief $\beta$. To determine $V_{i+1}$ from $V_i$, after choosing the spectrum sensing set $\mathcal K$ and observing $\vec{Y}$, the posterior belief is updated as 
    $\hat\beta=\hat{\mathbb{B}}(\beta,\mathcal{K},\vec{Y})$, the reward $R^*(\hat\beta)$ is accrued, the next prior belief is updated as $\beta_{next}=\mathbb B(\hat\beta;\vec\theta)$, so that $\gamma V_i(\beta_{next})$ is the discounted future value accrued; 
   the expectation is then taken with respect to $\vec{Y}$, whereas the maximization over $\mathcal K{\in}\mathcal A$
   yields the optimal spectrum sensing action that maximizes the expected reward plus discounted future value.}
Yet, this direct approach
is not applicable to our settings: 
1) the transition model $P_B(\vec{B}(t+1)|\vec{B}(t);\vec\theta)$ needed in \eqref{13} to update the
prior $\beta_{t+1}$
is unknown;
2)  as the number of subcarriers of interest increases, the number of states of the underlying MDP scales exponentially, resulting in a high-dimensional belief space and high complexity.
To address these two challenges:
\begin{itemize}[leftmargin=*]
    \item We incorporate an HMM EM estimator, i.e., the Baum-Welch algorithm, to learn the
    parameter vector $\vec{\theta}$ that defines the transition model $P_B(\vec{B}(t+1)|\vec{B}(t);\vec\theta)$, while concurrently solving for the sensing and access policy. This is developed in \secref{II.I}.
    \item We embed a low-complexity approximate PBVI algorithm known as PERSEUS \cite{WCL:13}, and embed it with fragmentation (into independent subsets of highly-correlated subcarriers), belief update simplification heuristics (Hamming distance state filters), and Monte-Carlo methods, developed in \secref{II.II}.
\end{itemize}
\vspace{-6mm}
\section{LESSA Algorithms}\label{II}
In this section, we propose a Baum-Welch algorithm to estimate
the parameter vector $\vec{\theta}$ that defines the Markov time-frequency correlation model.
In \secref{II.II}, we then use the estimated model in a randomized
PBVI algorithm known as PERSEUS \cite{WCL:13} to determine an \hlt{approximately optimal sensing policy}. Crucially, these algorithms are executed concurrently -- a key feature to enable adaptation  in non-stationary settings.
\vspace{-6mm}
\subsection{Occupancy Correlation Structure Estimation}\label{II.I}
Let $\tau$ be the learning period of the parameter estimation algorithm. Let $\mathbf{B}{=}[\vec{B}(t)]_{t{=}1}^{\tau}$ be the unknown sequence of states and $\mathbf{Y}{=}[\vec{Y}(t)]_{t{=}1}^{\tau}$ be the sequence of observations made at the CR's spectrum sensor from $t{=}1$ to $t{=}\tau$,
based on a generic spectrum sensing policy.
 We formulate the Maximum Likelihood Estimation (MLE) problem to estimate the vector $\vec{\theta}$ that defines the LU occupancy time-frequency correlation structure (detailed in \secref{I.II}) as
\begin{align}\label{19}
    \vec{\theta}^{*}=\argmax_{\vec{\theta}}\log{\Big(\sum_{\mathbf{B}}\mathbb{P}(\mathbf{B}|\vec{\theta})
    f(\mathbf{Y}|\mathbf{B},\boldsymbol{\mathcal K})\Big)}.
\end{align}
\hlt{Note that $f(\mathbf{Y}|\mathbf{B},\boldsymbol{\mathcal K})=\prod_{t=1}^\tau
f(\vec{Y}(t)|\vec{B}(t),\mathcal K_t)
$ is the observation pdf
conditional on the states, given by \eqref{8}, whereas
$\mathbb{P}(\mathbf{B}|\vec{\theta})$ can be expressed using the one-step parametric transition model as
$\mathbb{P}(\mathbf{B}|\vec{\theta})=
\mathbb{P}(\vec B(1))\prod_{t=2}^\tau
P_B(\vec{B}(t)|\vec{B}(t-1);\vec{\theta})
$.}
We solve the MLE problem using the Baum-Welch algorithm, an EM algorithm for HMMs \cite{Baum_1966} that estimates
$\vec{\theta}^{(i)}$ ($i$ denotes the iteration index) by iterating through the E- and M- steps: given $\vec{\theta}^{(i)}$, the E-step computes
\begin{align}
    & Q(\vec{\theta}|\vec{\theta}^{(i)})=\mathbb{E}_{\mathbf{B}|\mathbf{Y},\vec{\theta}^{(i)}}\left[\log{(
    \mathbb{P}(\mathbf{B}|\vec{\theta})
    f(\mathbf{Y}|\mathbf{B},\boldsymbol{\mathcal K})
    )}\right];
\end{align}
afterwards, the M-step 
 updates the estimate of $\vec\theta$ as
 \begin{align}\label{21}
    \vec{\theta}^{(i+1)}=\argmax_{\vec{\theta}}Q(\vec{\theta}|\vec{\theta}^{(i)}).
\end{align}
\hlt{Specifically, 
using the Markovian correlation structure,
and neglecting
additive terms independent of the optimization parameter $\vec\theta$ (indicated as $\propto$),
the E-step can be rewritten as
\begin{align}
\nonumber
&
Q(\vec{\theta}|\vec{\theta}^{(i)})
\propto\mathbb{E}_{\mathbf{B}|\mathbf{Y},\vec{\theta}^{(i)}}\Big[
\sum_{t=2}^\tau\log{(
P_B(\vec{B}(t)|\vec{B}(t{-}1),\vec{\theta})
    )}
    \Big],
            \\&
=\sum_{w\in\{0,1\}}[A_{w}^{(i)}(1)\log(q_{w})+A_{w}^{(i)}(0)\log(1-q_{w})]
\nonumber
\\&
+\sum_{u,v\in\{0,1\}}[B_{u,v}^{(i)}(1)\log(p_{u,v})+B_{u,v}^{(i)}(0)\log(1-p_{u,v})],
\end{align}
where we have defined
\begin{align}
&A_{w}^{(i)}(b)
\triangleq
\sum_{t=2}^\tau
\mathbb{P}(B_1(t)=b,B_1(t{-}1)=w|\mathbf{Y},\vec{\theta}^{(i)}),
\\&
B_{u,v}^{(i)}(b)
\triangleq
\sum_{t=2}^\tau\sum_{k=2}^K
\mathbb{P}(B_{k}(t)=b,B_{k-1}(t)=u,B_k(t{-}1)=v|\mathbf{Y},\vec{\theta}^{(i)}),
\end{align}
computed using the Forward-Backward algorithm \cite{Rabiner_1989}; the M-step  can then be optimized in closed-form as
$$
q_{w}^{(i+1)}=\frac{A_{w}^{(i)}(1)}{A_{w}^{(i)}(0)+A_{w}^{(i)}(1)},\ 
p_{u,v}^{(i+1)}=\frac{B_{u,v}^{(i)}(1)}{B_{u,v}^{(i)}(0)+B_{u,v}^{(i)}(1)}.
$$}

\label{TC:BW}
\hlt{The computational time complexity of the Baum-Welch algorithm is $O(\tau K \tilde{T})$ \cite{Rabiner_1989}, where $\tilde{T}$ is the number of iterations until convergence, which depends on the consistency of spectrum occupancy measurements, driven by our observation model and the CR's sensing limitations.}
\vspace{-5mm}
\subsection{The PERSEUS Algorithm with low-complexity approximations}\label{II.II}
LESSA solves for the spectrum sensing (and access, based on the reward maximization detailed in \secref{II.0}) policy in parallel with the parameter estimation algorithm, employing its published iterative transition model estimates, until both algorithms converge. We use  PERSEUS \cite{WCL:13} to devise an approximately optimal sensing policy, primarily motivated by the following rationale. Unlike the Exhaustive Enumeration  and the Witness algorithms in \cite{PUOccupancy:18}, PERSEUS does not perform the backup operation \eqref{18}
exhaustively on every reachable belief point; instead, it \emph{backs-up} only on a smaller subset of reachable belief points, and exploits this backup operation to improve the value of other reachable belief points, to achieve faster convergence; and instead of computing belief distances as in the PBVI algorithm \cite{PUOccupancy:17}, it computes  a \emph{finite} $U$-dimensional set of reachable beliefs $\tilde{\mathcal{B}}$ through an initial exploration phase, to balance computational complexity with accuracy (larger $U\Rightarrow$ improved accuracy but larger computational burden). 

Despite being an approximate POMDP method which alleviates the computational overhead associated with the backup operation, PERSEUS applied to the spectrum sensing problem still possesses computational intractability challenges due to the high-dimensional spectrum occupancy state $\vec{B}{\in}\{0,1\}^{K}$ and continuous observation space: the computational cost scales exponentially with the number of subcarriers $K$. 
To alleviate these challenges, we embed PERSEUS with three novel low-complexity enhancements.
\label{item:heuristics}
\begin{enumerate}[leftmargin=*]
\item \hlt{In updating the prior belief as in \eqref{13}, we avoid iterating over all possible $\vec B'\in\{0,1\}^K$
by considering only those state transitions that involve a Hamming distance of  at most $\delta{\in}\{1,2,{\dots},K\}$
subcarriers between two consecutive state vectors $\vec{B}(t)$ and $\vec{B}(t+1)$. This is practical because
spectrum occupancy of LUs typically varies slowly over time,
 slower than the processing dynamics of the POMDP agent.
 Let $\mathcal{B}_{\delta}(\vec{B}){\equiv}\{\vec{B}'{\in}\{0,1\}^K:\zeta(\vec{B},\vec{B}'){\leq}\delta\}$
be the set of spectrum occupancy states with Hamming distance ($\zeta$) at most $\delta$ from state $\vec{B}$. 
The next prior belief is then approximated as
\begin{align}
\label{approxprior}
\beta_{t+1}\approx\tilde{\mathbb{B}}(\hat{\beta}_{t},\vec\theta)\text{, where }\beta_{t+1}(\vec{B}){=}
\frac{
\sum_{\vec{B}'{\in}\mathcal{B}_{\delta}(\vec{B})}P_B(\vec{B}|\vec{B}';\vec{\theta})\hat{\beta}_t(\vec{B}')}
{
\sum_{\vec{B}''}
\sum_{\vec{B}'{\in}\mathcal{B}_{\delta}(\vec{B}'')}P_B(\vec{B}''|\vec{B}';\vec{\theta})\hat{\beta}_t(\vec{B}')
},
\end{align}
where the normalization ensures that $\beta_{t+1}$ is a probability distribution that sums to one.}

\item \hlt{We use a fragmentation technique to partition the set of $K$ subcarriers into independent smaller sets of adjacent subcarriers; we then run PERSEUS independently on each one of these fragments, concurrently and in parallel, by employing multi-threading capabilities in software frameworks.
For example, a radio environment with $18$ subcarriers  with a sensing constraint of $6$ subcarriers per time-slot is fragmented into $3$ independent fragments, each comprising $6$ subcarriers correlated by the occupancy behavior of the corresponding LUs, and 
a sensing constraint of $2$ subcarriers per time-slot.
This fragmentation is practical because in a radio environment with multiple LUs, each LU is typically restricted to a portion (a set of adjacent frequency bands) of the spectrum, either by design or by bureaucracy. We let $K'$ be the number of subcarriers, $\kappa'$ be the sensing restriction in each of these fragments.}

\item \hlt{We avoid the expectation with respect to the continuous observation vector in the backup operation \eqref{18} by using a Monte-Carlo method:
we generate $N$ independent realizations of $\vec Y|\vec B,\mathcal K$, based on the observation model \eqref{8};
we then compute the argument of the expectation with respect to each of these realizations, followed by a sample average.
As $N\to\infty$, this sample average converges to the conditional expectation in the backup operation.}
\end{enumerate}
\begin{algorithm}[t]
\small
\caption{Monte-Carlo Fragmented PERSEUS with Hamming distance state filters} 
\label{Alg. PERSEUS}
\begin{flushleft}
    \textbf{Input:} Parameter vector $\vec{\theta}$;
    $K'$ subcarriers, $\kappa'$ sensing restriction (for one fragment); \\Observation model as in \eqref{8}; Target accuracy $\epsilon>0$; Size of belief space $U$;\\
    \textbf{Output: } 
    Set of hyperplanes and spectrum sensing actions $\{(\vec\alpha^u,\mathcal K^u):u=1,\dots, U\}$ for spectrum fragment.
\end{flushleft}
\begin{algorithmic}[1]
    \vspace{-4mm}
    \State{Determine an $U$-dimensional set of \emph{reachable beliefs} $\tilde{\mathcal{B}}\equiv\{\beta_u:u=1,\dots,U\}$;} \Comment{\emph{Random exploration}}
    \State{{\bf Initialization:} Hyperplanes $\vec\alpha_0^u=\mathbf 0,\forall u$;
    actions $\mathcal K_0^u=\emptyset,\forall u$;
    iteration $i=-1$; $V_{0}(\beta)=0,V_{-1}(\beta)=-\infty;$}
    \While{$|V_{i+1}(\beta){-}V_{i}(\beta)|{>}\epsilon,{\exists}\beta{\in}\tilde{\mathcal{B}}$}
        \State{$i\leftarrow i+1$; \ \  $\tilde{\mathcal{U}}{\longleftarrow}\{1,\dots,U\}$;} \Comment{\emph{Start new iteration; init. hyperplanes indices}}
        \While{$\tilde{\mathcal{U}}{\neq}\{\cdot\}$} \Comment{\emph{Improve all points in $\tilde{\mathcal{U}}$}}
            \State{Pick $u$ randomly from $\tilde{\mathcal{U}}$, and $\beta_u$ from $\tilde{\mathcal B}$; $\tilde{\mathcal{U}}\leftarrow\tilde{\mathcal{U}}\setminus\{u\}$;}
            \State{For each $\mathcal K\in\mathcal A$, compute the hyperplane
                $\xi_{\mathcal{K}}^{u}$ with components}
            \State{$\xi_{\mathcal{K}}^{u}(\vec{B}){=}
            \frac{1}{N}\sum_{n=1}^N
            \Big[
                R(
                \vec{\phi}^{*}(\hat{\mathbb{B}}(\beta_{u},\mathcal{K},\vec{Y}_n))
                ,\vec{B})
                {+}\gamma\cdot\sum_{\vec{B}'}P_B(\vec{B}'|\vec{B};\vec\theta)\xi_{\mathcal{K},n}^{u}(\vec{B}')\Big]$, where}
            \State{$
    \vec{Y}_n\sim f(\cdot|\vec{B},\mathcal{K})
                $ i.i.d. over $n=1,\dots,N$ (Eq. \eqref{8}), and} \Comment{\emph{For Monte-Carlo expectation}}
            \State{$\xi_{\mathcal{K},n}^{u}{=}\!\!\underset{\vec\alpha_{i}^{u'},u'{\in}\{1,2,{\dots},|\tilde{\mathcal{B}}|\}}{\!\!\!\!\!\!\!\!\!\!\argmax}\!\!\!\!\!\!\!\!\!\!\!\!
                \langle\tilde{\mathbb{B}}(\hat{\mathbb{B}}(\beta_{u},\mathcal{K},\vec{Y}_n)),\vec\alpha_{i}^{u'}\rangle$;} \Comment{\emph{Future hyperplane with Hamming dist.-based belief update}}
            \State{$\mathcal{K}_{i+1}^{u}{=}\argmax_{\mathcal{K}{\in}\mathcal{A}}\langle\beta_{u},\xi_{\mathcal{K}}^{u}\rangle$,
                $\vec{\alpha}_{i+1}^{u}{=}\xi_{\mathcal K_{i+1}^{u}}^{u}$
                and $V_{i+1}(\beta_{u})=\langle\beta_{u},\vec{\alpha}_{i+1}^{u}\rangle$;} \Comment{\emph{Backup}}
            \For{$\forall u'{\in}\tilde{\mathcal{U}}$}
                \If{$\langle\beta_{u'},\vec{\alpha}_{i+1}^{u}\rangle{\geq}V_{i}(\beta_{u'})$}
                    \Comment{\emph{New hyperplane improves value of $\beta_{u'}$}}
                    \State{$V_{i+1}(\beta_{u'})=\langle\beta_{u'},\vec{\alpha}_{i+1}^{u}\rangle$, $\vec{\alpha}_{i+1}^{u'}=\vec{\alpha}_{i+1}^{u}$,
                    $\mathcal K_{i+1}^{u'}=\mathcal K_{i+1}^{u}$;}
                    \Comment{\emph{New value and action for $\beta_{u'}$}}
                    \State $\tilde{\mathcal{U}}{\longleftarrow}\tilde{\mathcal{U}}\setminus\{ u'\}$;
                    \Comment{\emph{Remove improved beliefs from $\tilde{\mathcal U}$}}
                \EndIf
            \EndFor
        \EndWhile
    \EndWhile
    \State{Return set of hyperplanes
    and associated spectrum sensing actions $\{(\vec\alpha^{*u},\mathcal K^{*u})\triangleq(\vec\alpha_{i}^u,\mathcal K_i^u):\ u=1,\dots, U\}$.}
    \end{algorithmic}
\end{algorithm}

The resulting \emph{Monte-Carlo Fragmented PERSEUS with Hamming distance-based belief update} is shown in Algorithm \ref{Alg. PERSEUS} (\hlt{run in parallel on each fragment of spectrum}), described below. 
The key idea of PERSEUS is to 
exploit the fact that
the value function $V_i$ generated by the backup operation \eqref{18} is Piece-Wise Linear Convex (PWLC) \cite{WCL:13},
hence it can be approximated by an 
 $U$-dimensional set of hyperplanes (each associated to a certain sensing action,
 Step 20) as
\begin{align}
\label{vapprox}
V(\beta_t)\approx
\max_{u{\in}\{1,2,{{\dots}},U\}}
\langle \beta_t,\vec{\alpha}^{*u}\rangle,
\end{align}
i.e., it is a PWLC function of $\beta_t$, where $\langle\beta,\vec{\alpha}\rangle{=}\sum_{\vec{B}}\beta(\vec{B})\vec{\alpha}(\vec{B})$ denotes inner product,
and $\{\vec\alpha^{*u}:\ u=1,\dots, U\}$ is the set of hyperplane vectors.
Letting $u_t$ be the maximizing index in \eqref{vapprox},
the approximately optimal spectrum sensing action is
$\mathcal K^{*u_t}$, associated with the maximizing hyperplane
$\vec{\alpha}^{*u_t}$. 
The goal of PERSEUS is to determine such a set of hyperplanes and associated actions. With these, the CR then operates as follows: with the prior belief $\beta_t$ at time-slot $t$,
it selects the spectrum sensing action $\mathcal K_t=\mathcal K^{*u_t}$; it then collects the observation $\vec Y(t)$, computes the posterior belief as 
$\hat\beta_t=\hat{\mathbb{B}}(\beta_t,\mathcal{K}_t,\vec{Y}(t))$, performs spectrum access as in \eqref{SAdecision}, and updates the next prior belief as $\hat\beta_{t+1}=\mathbb B(\hat\beta_t;\vec\theta)$, and so on.

To determine these hyperplanes and associated actions,
PERSEUS is first preceded by an initial phase of exploration
to determine a set of representative belief points $\tilde{\mathcal{B}}$, by allowing the CR to randomly interact with the radio environment (Step $1$). 
With $\tilde{\mathcal{B}}$ thus determined,
the hyperplanes and associated actions are computed
iteratively through Steps 3-19,
after initializing them in Step 2.
At iteration $i$,
starting from the entire set of beliefs, i.e., all indices $\{1,\dots,U\}$ (Step 4), the algorithm selects one belief at random, $\beta_u\in\tilde{\mathcal B}$ (Step 6).
For this belief point, it then performs the backup operation similarly to \eqref{18}, by
searching through all possible spectrum sensing actions (Step 7) to determine the optimal one that maximizes the value (Step 11). 
The value of a certain spectrum sensing action $\mathcal K$
is determined in Steps 8-10. To understand these steps, it is helpful to note that $\xi_{\mathcal{K}}^{u}(\vec{B})$ is the value accrued starting from state $\vec B$ and sensing action $\mathcal K$, so that $\langle \beta_u,\xi_{\mathcal{K}}^{u}\rangle$ done in Step 11 represents an expectation with respect to the belief $\beta_u$, i.e., the expected value of the sensing action $\mathcal K$ from belief $\beta_u$.
$\xi_{\mathcal{K}}^{u}(\vec{B})$ is comprised of two components: the instantaneous reward 
$R(\vec{\phi}^{*}(\hat{\mathbb{B}}(\beta_{u},\mathcal{K},\vec{Y}_n)),\vec{B})$ accrued in state $\vec B$, after 
observing $\vec Y_n$ and computing the spectrum access decision as in \eqref{SAdecision}; the discounted future reward 
$\sum_{\vec{B}'}P_B(\vec{B}'|\vec{B};\vec\theta)\xi_{\mathcal{K},\vec{Y}_n}^{u}(\vec{B}')$, obtained by averaging with respect to the one-step transition from $\vec B$ to $\vec B'$;
here, $\xi_{\mathcal{K},n}^{u}$ is the hyperplane associated to the future value function, determined in Step 10
as the one that maximizes the value function (with PWLC structure)
based on the next prior belief, computed via the Hamming distance-based method of \eqref{approxprior}.
Finally,  $\xi_{\mathcal{K}}^{u}(\vec{B})$  is obtained by taking an expectation with respect to the observations using the aforementioned Monte-Carlo method:
Step 9 generates $N$ independent realizations of $\vec Y$ given $\vec B$ and $\mathcal K$, so that 
$1/N\sum_n[\cdot]$ in Step 8 converges to $\mathbb E_{\vec Y|\vec B,\mathcal K}[\cdot]$ as $N\to\infty$.

Another key approximation of PERSEUS is to limit as much as possible the number of backup operations of Steps 7-11 by adding Steps 12-17:
here, the algorithm
scans through
the set of unimproved belief points indexed by $\tilde{\mathcal U}$; it then
checks whether the new hyperplane $\alpha_{i+1}^u$ determined in the backup operation improves the value of any remaining belief points; if it does for belief $\beta_{u'}$ (Step 13), then 
this new hyperplane (and its associated sensing action) becomes the relevant hyperplane (and the relevant sensing action) for this belief point (Step 14), and $u'$ is removed from $\tilde{\mathcal U}$ (Step 15), so that the backup operation is not done for $\beta_{u'}$ at iteration $i$.
These sequence of operations (random choice from $\tilde{\mathcal{U}}$, backup, check for improvement and removal) are performed iteratively until the set $\tilde{\mathcal{U}}$ is empty (Step $5$): this constitutes a single PERSEUS iteration. Multiple iterations are executed until the value iteration updates converge (Step $3$).

\label{TC:PERSEUS}
\hlt{The computational time complexity of 
Algorithm \ref{Alg. PERSEUS} is $O(\tilde{T}|\tilde{\mathcal{B}}|^{2}2^{2K'})$ \cite{WCL:13}, where $\tilde{T}$ denotes the number of iterations until convergence. Note here that incorporating Hamming distance state filters to alleviate the computational intractability inherent in PERSEUS belief update further mitigates the exponential dependence on the fragment size.}
\vspace{-3mm}
\section{Numerical Evaluations for LESSA}\label{III}
In this section, we evaluate numerically LESSA and compare it against state-of-the-art algorithms. The simulated radio environment constitutes $J_L{=}3$ LUs accessing a $2.88$MHz spectrum, discretized into $K{=}18$ subcarriers, each of bandwidth of $W{=}160$kHz, and one CR, as illustrated in \figref{fig: A.0}. The $3$ LUs access these $18$ subcarriers according to a time-frequency Markovian correlation structure defined in \secref{I.II}, with parameters $p_{0,0}=0.1,p_{0,1}=p_{1,0}=0.3,p_{1,1}=0.7$, $q_{0}=0.3,q_{1}=0.8$. \hlt{LUs' and CR's transmitters and receivers are deployed randomly across an operational region: the three LUs' transmitters are placed at positions $(-225,200)$m, $(225,200)$m and $(0,-300)$m, at a height of $40$m; the corresponding LU receivers are stationary nodes, placed randomly within a circular radius of $200$m from the respective LU's transmitter. The CR's transmitter is located at position $(0,0)$m at a height of 20m; the corresponding receiver is at ground level, and moves along a randomly generated trajectory within a radius of $100$m from the CR's transmitter. As described below, we have also facilitated rate adaptation at the CR based on the perceived SINR, which may vary with time and channel indices. The CR is capable of sensing $\kappa{=}6$ subcarriers per time-slot.}

\hlt{For each transmitter (Tx, LU or CR) and receiver  (Rx; the intended one, or unintended, thus experiencing interference) pairs, we model the channel on the $k$th subcarrier as $\hbar_k{=}\sqrt{\psi}\omega_k$, where $\psi$ and $\omega_k{\in}\mathbb C$ encapsulate the large- and small-scale channel variations, respectively -- with $\mathbb{E}[|\omega_k|^{2}]{=}1$. Given the angle of elevation between the Tx/Rx pair under consideration, $\chi{\in}(0,\frac{\pi}{2}]$, we generate the line-of-sight (LoS) condition with probability $P_{\text{LoS}}(\chi){=}\frac{1}{1{+}z_{1}e^{-z_{2}(\chi{-}z_{1})}}$  \cite{Tse}, so that the non-LoS (NLoS) condition occurs with probability  $P_{\text{NLoS}}(\chi){=}1-P_{\text{LoS}}(\chi)$, where $z_{1}, z_{2}$ are parameters specific to the propagation environment. Given the LoS or NLoS condition, and the distance $d$ between Tx and Rx, we then generate the large- and small-scale channel conditions as follows. If the channel is LoS: the large-scale fading coefficient $\psi$ is modeled as $\psi_{\text{LoS}}(d){=}\psi_{0}d^{-\mu_L}$, where $\psi_{0}$ is the reference pathloss at a distance of $1$m from the Tx and $\mu_L{\geq}2$ is the LoS pathloss exponent; the small-scale fading coefficient $\omega_k$ is modeled as a Rician with K-factor $\mathbb{K}(\chi){=}f_{1}e^{f_{2}\chi}$, where $f_{1}, f_{2}$ are parameters specific to the propagation environment. Conversely, if the channel is  NLoS: the large-scale fading coefficient $\psi$ is modeled as $\psi_{\text{NLoS}}(d){=}\iota\psi_{0}d^{-\mu_N}$, where $\iota{\in}(0, 1]$ is the additional NLoS attenuation and $\mu_N{\geq}\mu_L$ is the NLoS pathloss exponent; the small-scale fading coefficient $\omega_k$ is modeled as Rayleigh distributed (i.e., Rician with K-factor $\mathbb{K}(\chi){=}0$). Throughout the simulation, we use $\mu_{L}{=}2.0,\ \mu_{N}{=}2.8,\ \iota{=}0.2,\ W{=}160\ \text{kHz},\ f_{1}{=}1.0,\ f_{2}{=}0.0512,\ z_{1}{=}9.12, \text{ and } z_{2}{=}0.16$ \cite{Urban_Model}.}

\hlt{With this channel model, the SINR between the $i$th transmitter (LU or CR) and its intended receiver on subcarrier $k$ is given by
$$
\text{SINR}_{k,i}=\frac{|\hbar_{k,i,i}|^2P_{k,i}}{\sigma_{V}^{2}+\sum_{j\neq i}|\hbar_{k,j,i}|^2P_{k,j}},
$$
where $\hbar_{k,i,i}$ is the channel
between the $i$th transmitter and its indented receiver,
$P_{k,i}$ is the transmission power; $\sigma_{V}^{2}$ is the noise power
and $\sum_{j\neq i}|\hbar_{k,j,i}|^2P_{k,j}$ is the contribution from the active interferers: 
$\hbar_{k,j,i}$ is the channel between the $j$th interfering transmitter and the $i$th receiver, with transmission power $P_{k,j}$ ($P_{k,j}=0$ if such transmitter is idle on subcarrier $k$).
We can then define the link capacity for such transmitter-receiver pair on subcarrier $k$ as
$$C_{k,i}=W\log_2(1+\text{SINR}_{k,i}).$$
We assume that the LUs transmit with fixed transmission rate of
$\Phi_{\text{LU}}{=}0.9$Mbps (per subcarrier), and thus incur outage if $\text{SINR}<2^{\Phi_{\text{LU}}/W}-1$.
We now detail the choice of the transmission rate at the CR, $\Phi_{\text{CR}}$.
Assuming that the large-scale fading coefficient $\psi$ and the K-factor $\mathbb{K}$ are known throughout the simulation period (since they represent large-scale parameters, they can be reliably estimated), the outage probability for an interference-free link is given by \cite{Tse}
\begin{align}\nonumber
    P_{\text{out}}(\Phi_{\text{CR}},\psi,\mathbb{K})&
    {=}\mathbb{P}(C_{k,i}{<}\Phi_{\text{CR}}{|}\psi,\mathbb{K}){=}
    1{-}Q_{1}\left(\sqrt{2\mathbb{K}}{,}\sqrt{\frac{2(\mathbb{K}{+}1)\sigma_{V}^{2}
    }{\psi P_{T}}}(2^{\frac{\Phi_{\text{CR}}}{W}}{-}1)\right),
\end{align}
where $Q_{1}$ denotes the standard Marcum Q-function. 
Herein, we select the rate for the CR as the one that maximizes the expected throughput $\Phi_{\text{CR}}\cdot[1-P_{\text{out}}(\Phi_{\text{CR}},\psi,\mathbb{K})]$,
  obtained efficiently via a bisection method \cite{Convex}. 
}

To evaluate the performance, we define the average CR's throughput over $T$ time-slots as
\begin{align}\label{30}
    C^{\text{CR}}=\frac{1}{T}\sum_{t=1}^{T}\sum_{k=1}^{K}\Phi_{\text{CR},k}(t)\phi_k(t)\mathcal I\left[\text{SINR}_{\text{CR},k}(t) \geq 2^{\frac{\Phi_{\text{CR},k}(t)}{W}}-1\right],
\end{align}
where $\Phi_{\text{CR},k}(t)$ is the rate used by the CR on subcarrier $k$ in time-slot $t$, based on the rate adaptation scheme described earlier, and $\vec\phi(t)$ denotes the spectrum access decision in time-slot $t$;
we define
the LUs' throughput over the same $T$ time-slots, normalized by the number of transmissions as
\begin{align}\label{31}
    C^{\text{LUs}}=\frac{\sum_{t=1}^{T}\sum_{k=1}^{K}\Phi_{\text{LU}}B_{k}(t)\mathcal I\left[\text{SINR}_{\text{LU},k}(t) \geq 2^{\frac{\Phi_{\text{LU}}}{W}}-1\right]}{\sum_{t=1}^{T}\sum_{k=1}^{K}B_{k}(t)},
\end{align}
where $\text{SINR}_{\text{LU},k}(t)$ is the SINR of the occupying LU,
computed based on the channel realizations as described earlier.

\begin{figure} [t]
     \begin{subfigure}{0.5\linewidth}
         \centering
         \includegraphics[width=0.9\linewidth]{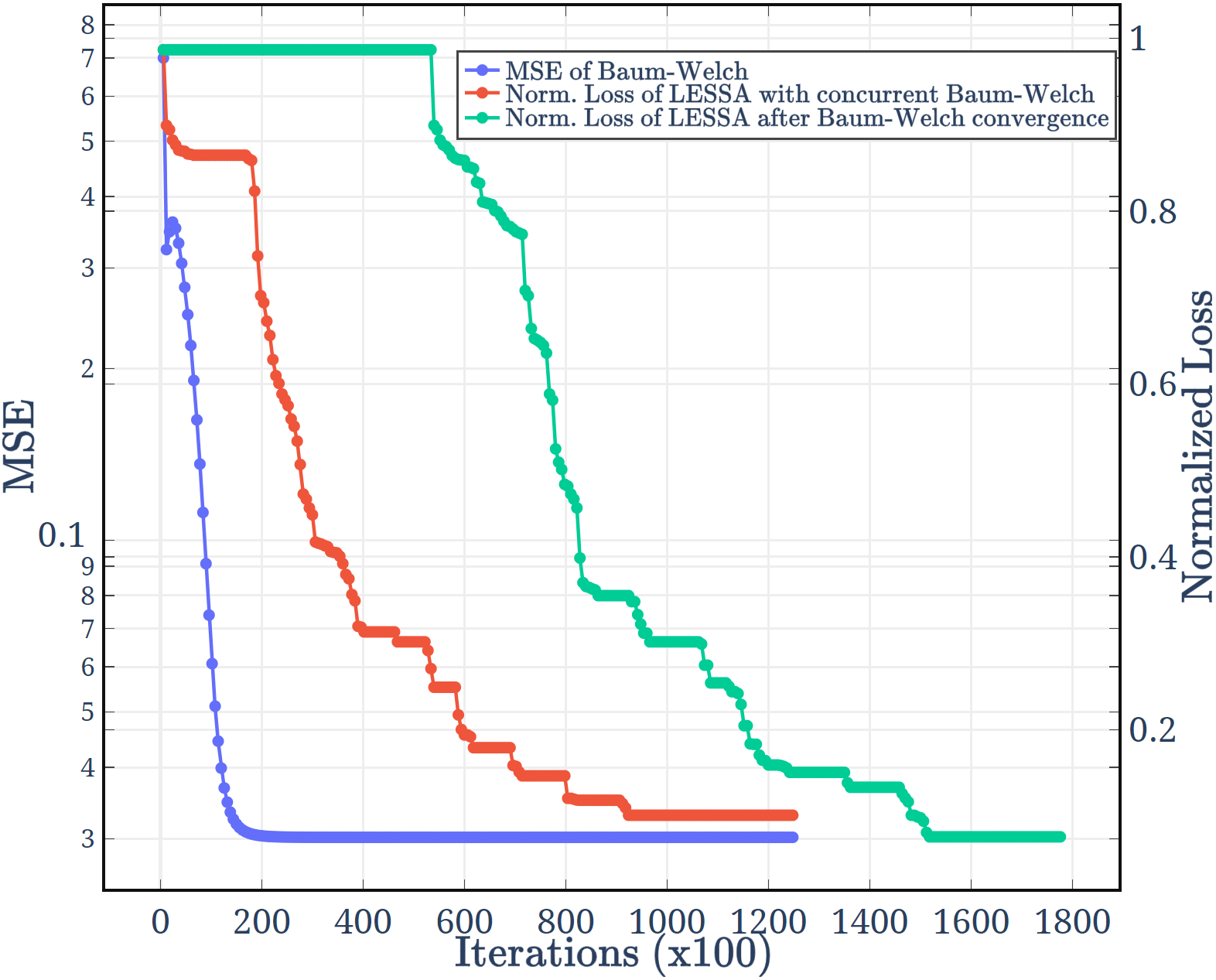}
         \caption{Convergence visualization}
         \label{fig:convergence}
     \end{subfigure}
     \begin{subfigure}{0.5\linewidth}
         \centering
         \includegraphics[width=0.9\linewidth]{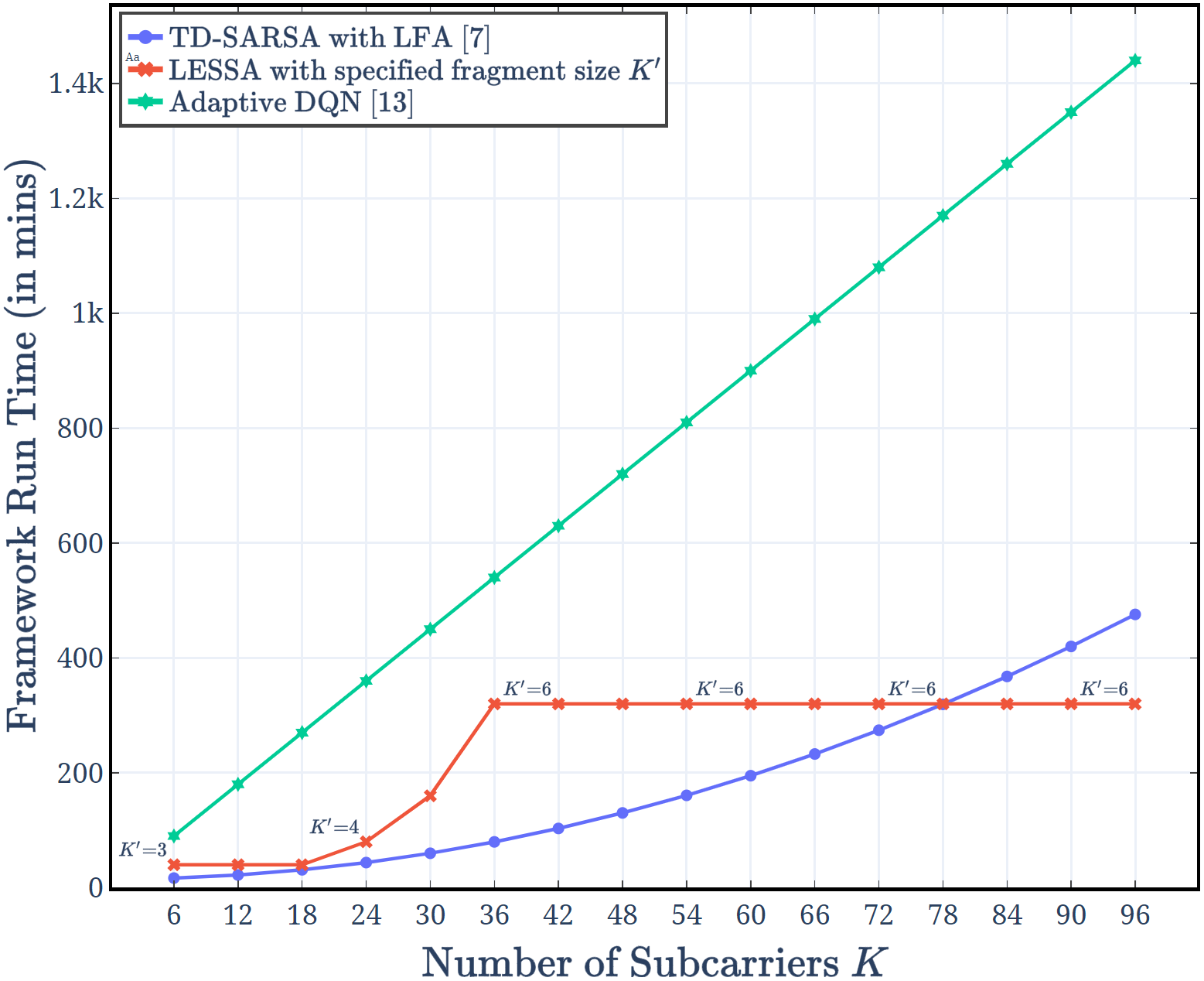}
         \caption{Computational Bench-marking}
         \label{fig:benchmark}
     \end{subfigure}
     \vspace{-4mm}
     \caption{MSE convergence of the Baum-Welch algorithm to estimate $\vec{\theta}$, and normalized loss incurred by Algorithm~\ref{Alg. PERSEUS} (a); \hlt{computational time complexity bench-marking of LESSA against state-of-the-art (b)}.}
     \label{fig:convergence_benchmark}
     \vspace{-10mm}
\end{figure}

In \figref{fig:convergence}, we plot the Mean Square Error (MSE) of the Baum-Welch algorithm versus the number of EM steps ($i$), $\Vert\vec{\theta}-\hat{\vec{\theta}}^{(i)}\Vert_{2}^{2}$, averaged over multiple independent realizations. The parameters are initialized as $p_{u,v}{=}0.5,{\forall}u,v{\in}\{0,1\}$ and $q_{w}{=}0.5,w{\in}\{0,1\}$. We observe that the MSE decreases over time, as the estimation progresses through the E- and M-steps \cite{Rabiner_1989}, and converges to an MSE value of $0.03$ after $\sim 5\times 10^4$ iterations: this corresponds to an observation and estimation period of $160$s, considering a typical time-slot duration of $3$ms.

\label{MSEconvergence}
In \figref{fig:convergence}, we also plot the convergence of the PERSEUS algorithm, using a discount factor of $\gamma{=}0.9$ and a termination threshold of $\epsilon{=}10^{-5}$, $\lambda{=}1$, for both cases in which it is run concurrently with the Baum-Welch algorithm (red curve) or after its convergence (green curve).
To evaluate convergence, we use a normalized loss metric, defined as the loss of utility (defined in \eqref{12a}) with respect to an oracle that performs spectrum access based on knowledge of the current occupancy state $\vec B(t)$, defined mathematically as
$$
1-\frac{
R(\vec{\phi}^{*}(\hat{\beta}_{t}),\vec{B}(t))}{\max_{\vec{\phi}\in\{0,1\}^K}R(\vec{\phi},\vec{B}(t))},
$$
averaged out over time and multiple realizations,
where $\vec{\phi}^{*}(\hat{\beta}_{t})$ is the spectrum access decision defined in \eqref{SAdecision}, based on the posterior belief $\hat\beta_t$ generated in our POMDP formulation,
and $\max_{\vec{\phi}\in\{0,1\}^K}R(\vec{\phi},\vec{B}(t))$ is the Oracle reward, which uses knowledge of $\vec{B}(t)$.
\hlt{We find that the 
 concurrent method converges in approximately half the time of the non-concurrent one. Remarkably, the normalized loss
 is around $5\%$ after convergence, i.e.,
LESSA performs on par with an Oracle
with knowledge of the current spectrum occupancy
 --
a significant result considering the lack of \textit{a priori} knowledge of the underlying Markov transition model and the noisy and constrained spectrum sensing environment.
In addition, since the Oracle performs better than the \emph{optimal POMDP} policy, this result demonstrates that
solving for the \emph{optimal POMDP} policy (a computationally intractable task) does not yield a tangible boost in spectrum white-space detection, thereby legitimizing the validity of the approximate POMDP approach proposed in this paper.}

\label{TC:simres}
\hlt{\figref{fig:benchmark} plots the run time of LESSA as a function of the number of subcarriers,  against TD-SARSA with LFA \cite{WCL:5} and Adaptive DQN \cite{WCL:DQN}. The algorithms are run on a $2\times$ $12$-core Intel Xeon Gold $6126$ @ $2.6$GHz compute node with $192$GB RAM \cite{Duplyakin+:ATC19}: as the number of subcarriers increases, our solution scales better yielding a more computationally tractable performance relative to the other two, owing to fragmentation and belief update simplification heuristics. Interestingly, for $K\geq 36$ with fixed fragment size of $K'=6$, the run time of LESSA flattens out: this is because LESSA is carried out in parallel across all $6$-subcarrier fragments.}

\begin{figure} [t]
    \centerline{
    \includegraphics[width=0.6\linewidth]{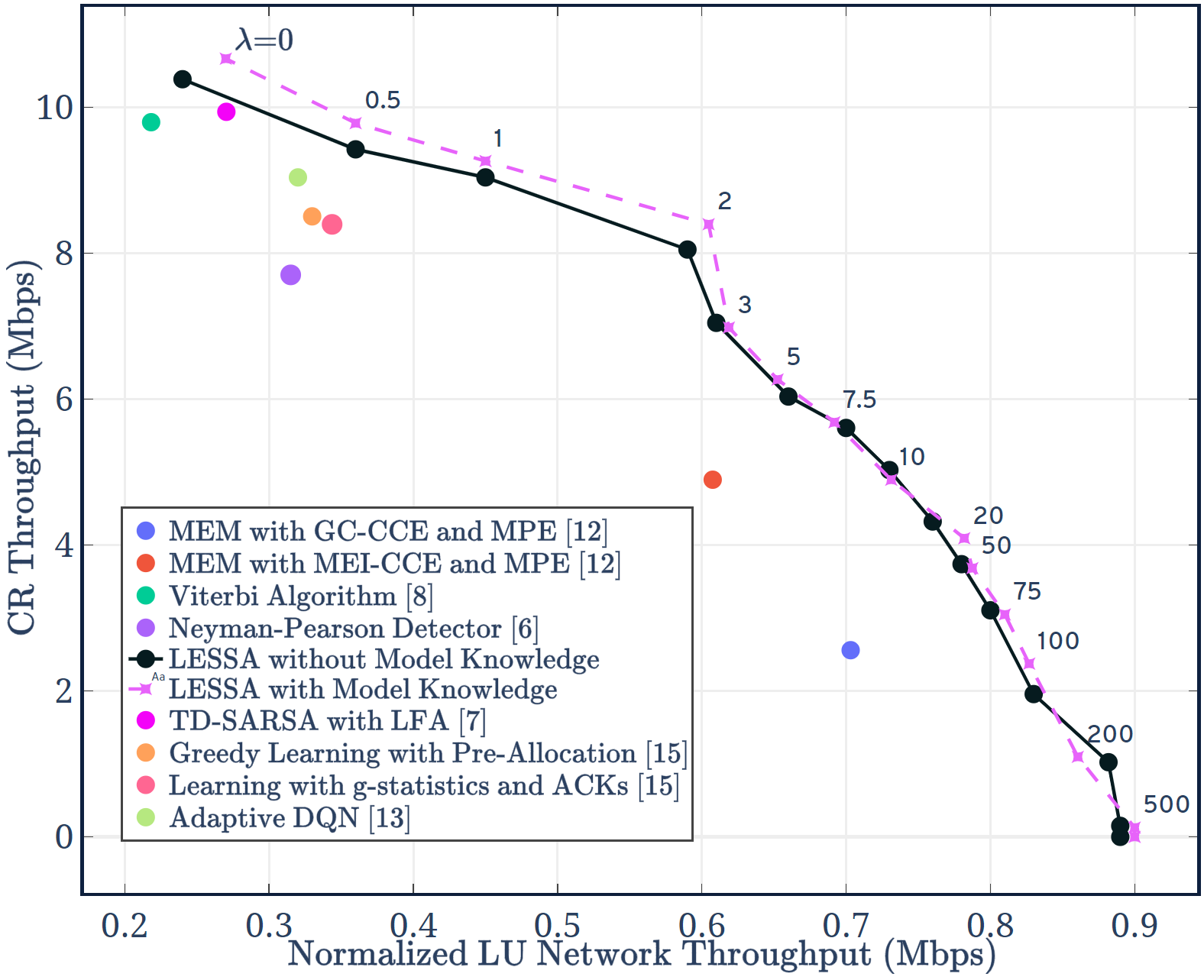}}
    \vspace{-4mm}
    \caption{Evaluation of CR and LU network throughputs for different values of $\lambda$ \hlt{with rate adaptation at the CR} and with/without correlation model foreknowledge -- along with comparisons with the state-of-the-art.}
    \vspace{-10mm}
    \label{Fig. 4}
\end{figure}

\begin{figure} [t]
     \begin{subfigure}{0.5\linewidth}
         \centering
         \includegraphics[width=0.9\linewidth]{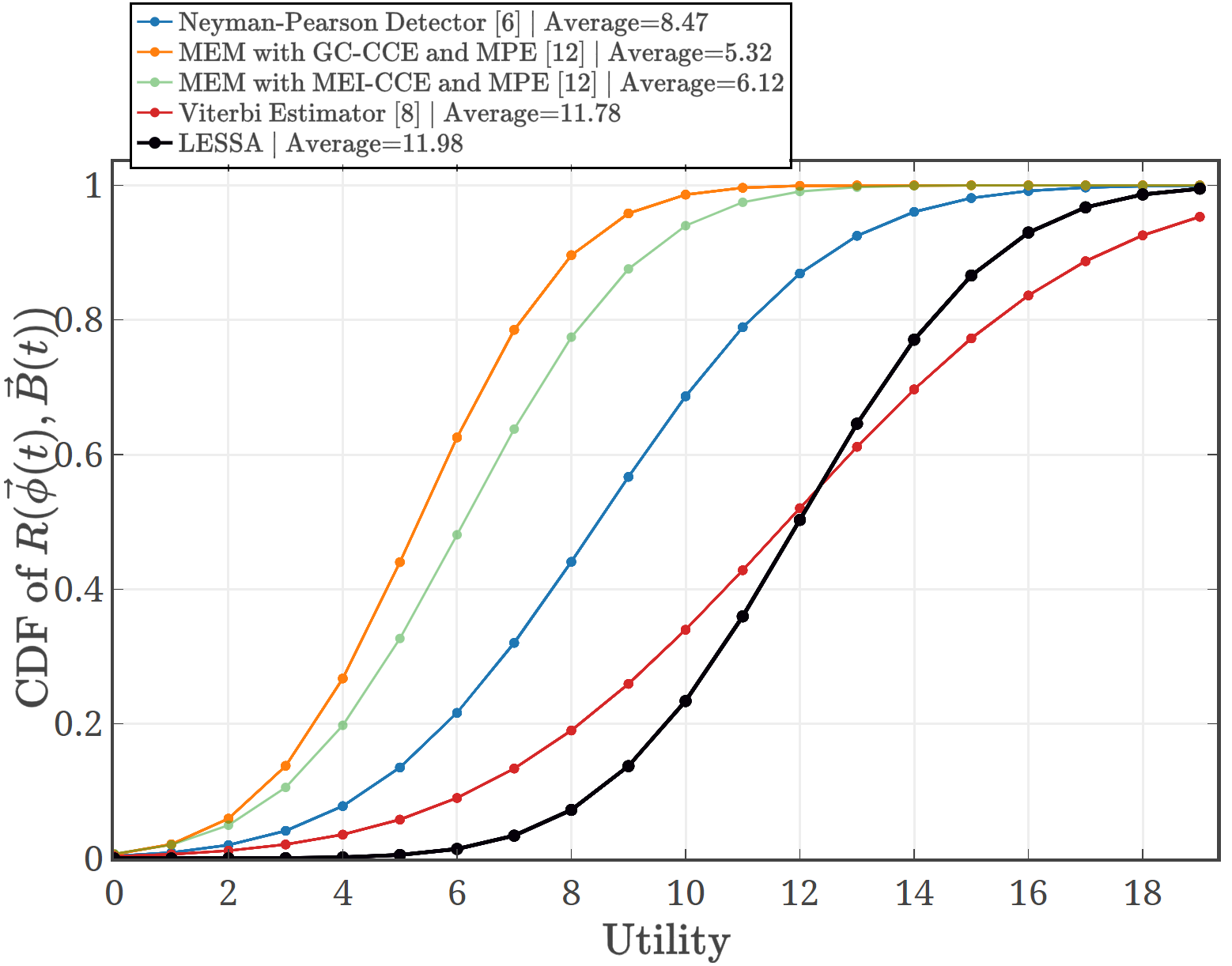}
         \caption{CDF of utility}
         \label{fig:utilities_single}
     \end{subfigure}
     \begin{subfigure}{0.5\linewidth}
         \centering
         \includegraphics[width=0.8\linewidth]{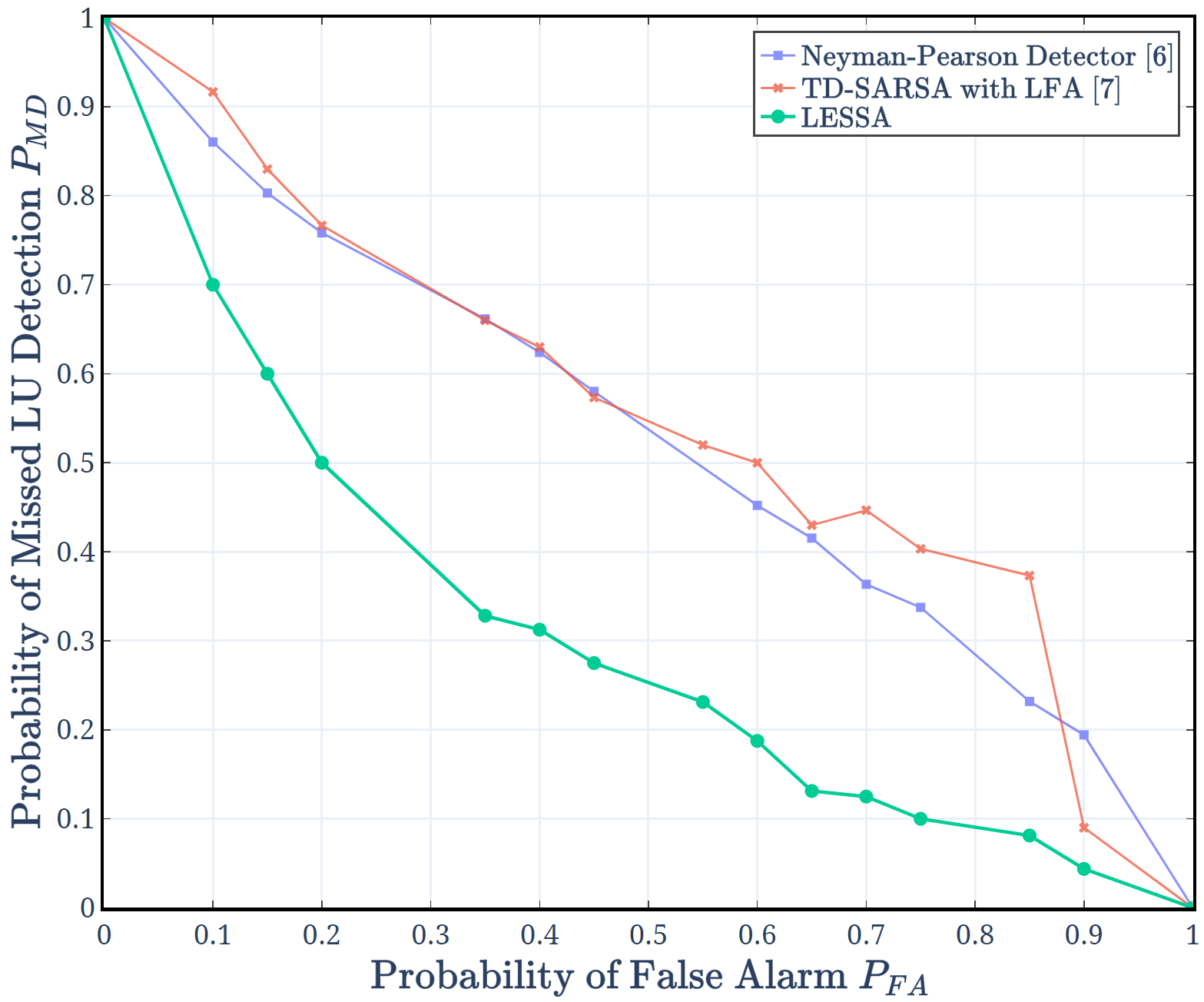}
         \caption{Receiver Operating Characteristics}
         \label{fig:roc}
     \end{subfigure}
     \vspace{-4mm}
     \caption{CDF of the utility $R(\vec{\phi}(t),\vec{B}(t))$ (see \eqref{12a}), and comparison with the state-of-the-art (a); \hlt{The receiver operating characteristics of LESSA versus those of a Neyman-Pearson Detector \cite{WCL:11} and TD-SARSA with LFA \cite{WCL:5}}.}
     \label{fig:utilities_and_roc}
     \vspace{-10mm}
\end{figure}

In \figref{Fig. 4}, we compare the performance of LESSA with state-of-the-art algorithms, in terms of the CR's network throughput achieved vis-à-vis LUs throughput, as defined in \eqref{30} and \eqref{31}, respectively.
We note that LESSA covers the entire trade-off region by varying the penalty weight $\lambda$ in the reward metric of the POMDP.
Generally, \figref{Fig. 4} depicts a trend of increasing CR throughput and
decreasing LUs' throughput (due to increased CR's interference), as the penalty $\lambda$ is decreased. Therefore, our framework provides a crucial practical tool in CR MAC design: the ability to tune the trade-off between the throughput obtained by the CR and the interference caused by it to LU transmissions in the network. 
In contrast, the other state-of-the-art algorithms do not offer such capability, hence they attain a single point in the trade-off region. 
By comparing LESSA with unknown model (learned with the Baum-Welch algorithm) and LESSA with model known \textit{a priori}, we observe that prior knowledge of the model offers a meagre $3.75$\% improvement in CR throughput for a given LU network throughput, compared to the proposed online concurrent model estimation and policy solver strategy -- a testament to the accuracy of our estimator. Comparing LESSA with other state-of-the-art schemes, we observe consistent improvement in performance across the entire trade-off region.
Specifically:\footnote{Unless otherwise stated, all these algorithms have a spectrum sensing restriction of $\kappa=6$; implementation details can be found in the respective references; the percentage improvements are all referred to
improvements in CR's throughput provided by LESSA under the same LUs' throughput achieved by the state-of-the-art scheme under consideration.}
\begin{itemize}[leftmargin=*]
    \item 
    Minimum Entropy Merging (MEM) with Greedy Clustering (GC) based Channel Correlation Estimation (CCE) and Markov Process Estimation (MPE) \cite{WCL:7}: Correlation Threshold $0.7$; our solution offers a $115$\% improvement over this strategy;
    \item MEM with Minimum Entropy Increment (MEI) based CCE and MPE \cite{WCL:7}: Correlation Threshold $0.7$; our solution achieves $40$\% better performance over this strategy. In both GC and MEI implementations, our improved performance is owed to the POMDP formulation instead of an offline correlation-coefficient based clustering scheme employed in their work;
    \item Imperfect HMM-MAP State Estimation \cite{WCL:6}: it uses the Viterbi algorithm; although \cite{WCL:6} assumes time-only correlation and no sensing restrictions, we extend it to our setup to include \textit{a priori} knowledge of the time-frequency correlation model, and sensing restriction of $\kappa=6$; our solution attains a $6$\% boost over this strategy
    owing to an adaptive sensing policy exploited in our solution as opposed to a fixed one in their system, despite the lack of a-priori model knowledge;
    \item Neyman-Pearson Detection \cite{WCL:11}: it assumes
    time-frequency independence, and no channel sensing restrictions, an AND fusion rule across $300$ samples, and threshold determination via a false alarm probability of $30$\%; our solution offers a $26$\% enhancement over this strategy, by leveraging the time-frequency LU occupancy dynamics in our framework in contrast to the independence assumptions made in their model, despite the sensing restrictions of our model;
    \item \hlt{Temporal Difference Learning via SARSA with Linear Function Approximation (LFA) \cite{WCL:5}: we use a belief update heuristic constant of $0.9$, a discount factor of $0.9$, a fixed exploration factor of $0.01$, and a raw false alarm probability of $5$\%; our solution exhibits a $3$\% superior performance over this strategy, owing to our additional exploitation of LU occupancy correlation across frequency as opposed to a purely temporal strategy employed in their model;}
    \item Greedy Learning under Pre-Allocation \cite{WCL:MIT}:
    it uses a time-varying exploration factor, and assumes independence across both time and frequency;
    \item Learning with g-statistics and ACKs \cite{WCL:MIT}: our solution achieves a $13$\% boost in performance over this strategy. In both implementations of \cite{WCL:MIT}, LESSA offers enhancements by leveraging the time-frequency correlation structure in LUs' spectrum occupancy, in contrast to the independence assumptions made in their model;
    \item \hlt{Adaptive Deep Q-Networks \cite{WCL:DQN}:
     with Experiential Replay (Memory Size $10^{6}$), $2048$ input neurons, $4096$ neurons with ReLU activation functions in each of the $2$ hidden layers of the neural network, a MSE cost function with an Adam optimizer, a fixed exploration factor of $0.1$, a learning rate of $10^{-4}$, a batch Size of $32$; our solution offers a $9$\% improvement over this strategy, owing to a parametric time-frequency correlation model that captures the correlation in LU occupancy,
     as opposed to a model-free policy optimization scheme used in their system.}
\end{itemize}
In \figref{fig:utilities_single}, we plot the Cumulative Distribution Function (CDF) of the reward $R(\vec{\phi}(t), \vec{B}(t))$, evaluated according to \eqref{12}. We find that LEMMA achieves an average utility of
$11.98$ per time-step, $125$\% higher than that achieved by the MEM with GC based CCE and MPE algorithm from \cite{WCL:7}, $96$\% higher than that achieved by the MEM with MEI based CCE and MPE algorithm from \cite{WCL:7}, and $42$\% higher than that attained by the Neyman-Pearson Detector \cite{WCL:11} (despite the latter having no sensing restrictions). Compared to the Viterbi algorithm \cite{WCL:6} (${=}11.78$), our scheme achieves $2$\% higher utility (${=}11.98$), thanks to an adaptive sensing strategy.

\hlt{Finally, in \figref{fig:roc}, we illustrate the receiver operating characteristic, i.e., the trade-off between false alarm probability ($P_{FA}{=}\mathbb{P}(\phi_{k}(t){=}0{|}B_{k}(t){=}0$, i.e., the subcarrier is incorrectly detected as occupied, and it is therefore not used by the CR) versus missed detection probability ($P_{MD}{=}\mathbb{P}(\phi_{k}(t){=}1{|}B_{k}(t){=}1$, i.e., the subcarrier is incorrectly detected as idle, and it is therefore used by the CR), averaged out over time-slots and subcarriers, for LESSA ($\kappa{=}6$), and state-of-the-art TD-SARSA with LFA ($\kappa{=}6$) \cite{WCL:5} and Neyman-Pearson Detection (no sensing restrictions) \cite{WCL:11}. Although the algorithms in \cite{WCL:11} and \cite{WCL:5} do not inherently possess means to regulate the trade-off between CR throughput and LU network interference, for the sake of this evaluation, we have modified these algorithms to enable such a mechanism: since \cite{WCL:11} involves energy detection, false alarm constraints can be brought in by modifying the detection threshold; and we add this regulation to the observation and access decision reliability control logic in \cite{WCL:5}. In line with the results of Fig. \ref{Fig. 4}, we observe that LESSA consistently achieves lower $P_{MD}$ (hence lower interference to LUs) across all ranges of $P_{FA}$. For instance, at a false alarm probability target of $P_{FA}{=}0.7$, LESSA achieves $P_{MD}{=}0.125$, as opposed to $0.36$ by the Neyman-Pearson Detector (despite no sensing restrictions) \cite{WCL:11} and $0.45$ by TD-SARSA with LFA \cite{WCL:5}.}
\vspace{-3mm}
\section{Multi-Agent LESSA: Algorithms, Implementation and Evaluation}\label{Z}
In this section, we extend LESSA to both centralized and distributed multi-agent deployments (MA-LESSA). With respect to distributed deployments, we propose novel neighbor discovery and channel access rank allocation schemes that are embedded into LESSA -- as depicted in \figref{fig: A.add-1} -- along with steps for sharing localized observations and access decisions with neighbors, as well as the corresponding data aggregation. The proposed neighbor discovery scheme involves an adapted version of the Multi-band Directional Neighbor Discovery (MDND) algorithm outlined in \cite{MDND}, with RSSI-thresholding based modifications; furthermore, inspired by quorum-based voting heuristics for cluster fallback mechanisms in Message Oriented Middleware frameworks  \cite{AMQ}, we detail an access order determination scheme for cooperative access among CRs in the ensemble. As a part of our centralized deployment analysis, we retrofit MA-LESSA into the Purdue BAM! Wireless CR network \cite{BAM}, with a gateway CR handling flow scheduling to individual nodes in the network, making multi-hop routing decisions, collaborating with other competitor networks, and ensuring that Quality of Service (QoS) mandates are being met throughout the scenario emulation period on the DARPA SC$2$ Colosseum \cite{DARPA:SC2}.

{\bf Neighbor Discovery via RSSI-Thresholding:} The MDND scheme detailed in \cite{MDND} employs the $2.4$GHz Wi-Fi band as a control channel for neighbor discovery in $60$GHz data environments. Adapting this scheme to multi-agent CR deployments, we designate the band-edges as the control channel on which CRs communicate and coordinate their neighborhoods for channel access order allocation and data aggregation. At a pre-determined schedule, each CR broadcasts its control frames (with a frame header and node identifier) over the control channel, and upon receiving control messages from all its surrounding nodes, each CR checks if the expected RSSI of the radio signals corresponding to a certain node is above a threshold $\text{RSSI}_\text{th}$: if that is the case, it adds that node’s identifier to its list of neighbors. \hlt{The computational time complexity of the RSSI thresholding scheme for neighbor discovery is $O(J_C^{2})$.}
\label{TC:RSSIthres}

{\bf Channel Access Rank Allocation via Quorum-based Preferential Ballot:} With a similar control channel strategy for channel access rank allocation, inspired by cluster fallback schemes in queueing environments in enterprise software systems \cite{AMQ}, we employ a quorum-based preferential ballot scheme to determine the order in which the \emph{estimated-idle} subcarriers are accessed by the CRs in the network. This procedure starts only after a quorum has been achieved, i.e., the number of neighbors identified by an CR should be equal to or exceed a node-specific pre-defined number. Over the control channel, each CR exchanges a ranked list of its neighbors in the decreasing order of their respective RSSIs, with itself being on the list at position-1 (ties are broken via uniform random choice). Upon receiving an \emph{RSSI-ranked} list from one of its neighbors, each CR assigns points to each ranked position, with higher ranks getting larger point values, and re-broadcasts an \emph{aggregated-ranked} list of neighbors (with itself being on the list) with the ranking based on the point-values aggregated across all the ranked lists received from its neighbors (ties are broken via uniform random choice). If the \emph{aggregated-ranked} lists received from its neighbors matches the one at the CR for a pre-specified consecutive period of time, a consensus has been reached, so that the channel access order is determined by this \emph{harmonized-aggregated-ranked} list. If the \emph{aggregated-ranked} lists received from its neighbors differ from the one at the CR, then  the CR repeats the re-ranking of these list members based on their new aggregated point-values and broadcasts the new \emph{aggregated-ranked} list to its neighbors over the control channel. This repetitive process continues until a consensus is reached. \hlt{The computational time-complexity of the quorum-based preferential ballot scheme for channel access rank allocation is $O(\tilde{T}{J}_{C}^{2})$ \cite{WCL:MIT}, where $\tilde{T}$ corresponds to the number of iterations involved until a consensus is reached -- which depends on the mobility patterns of these nodes along with the temporal evolution of the peer-to-peer link qualities.}
\label{TC:prefball}
\vspace{-5mm}
\subsection{Distributed MA-LESSA}
Here, we evaluate the operational capabilities of MA-LESSA in distributed settings. As illustrated in \figref{fig: A.0}, operating under the same signal and observation models as in \secref{t}, consider a network of $3$ LUs operating in an 18-subcarrier radio environment, with their occupancy behaviors governed by Markovian time-frequency correlation structure ($\vec{\theta})$, and $12$ CRs intelligently trying to access white-spaces in the spectrum (cooperatively \cite{WCL:5} or opportunistically \cite{WCL:MIT}), with an added restriction of being able to sense only $1$ subcarrier per CR per time-slot.

\begin{figure} [t]
     \begin{subfigure}{0.48\linewidth}
         \centering
         \includegraphics[width=0.9\linewidth]{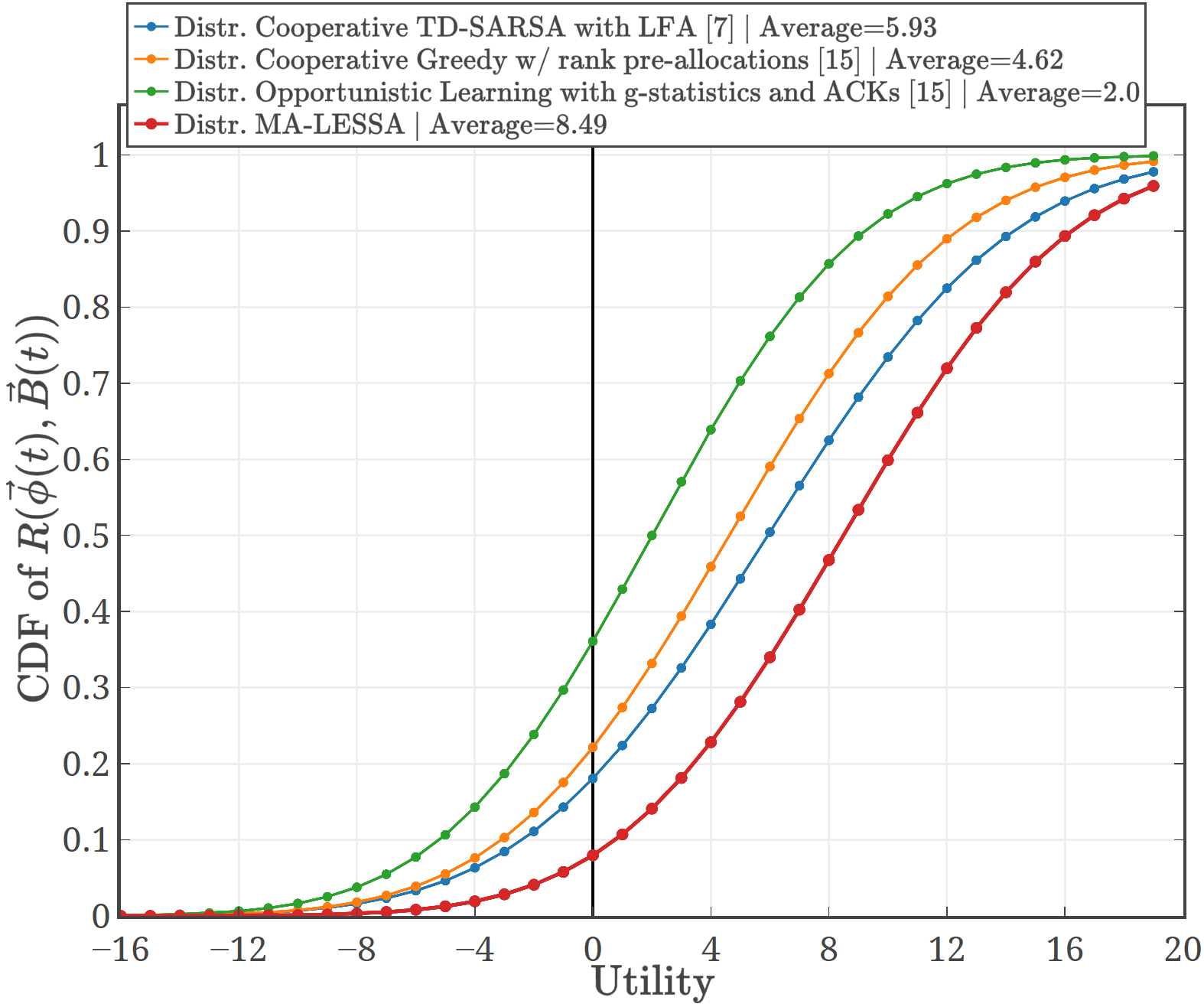}
         \caption{Distributed MA-LESSA Episodic Utilities}
         \label{fig: Z. 2}
     \end{subfigure}
     \begin{subfigure}{0.52\linewidth}
         \centering
         \includegraphics[width=0.9\linewidth]{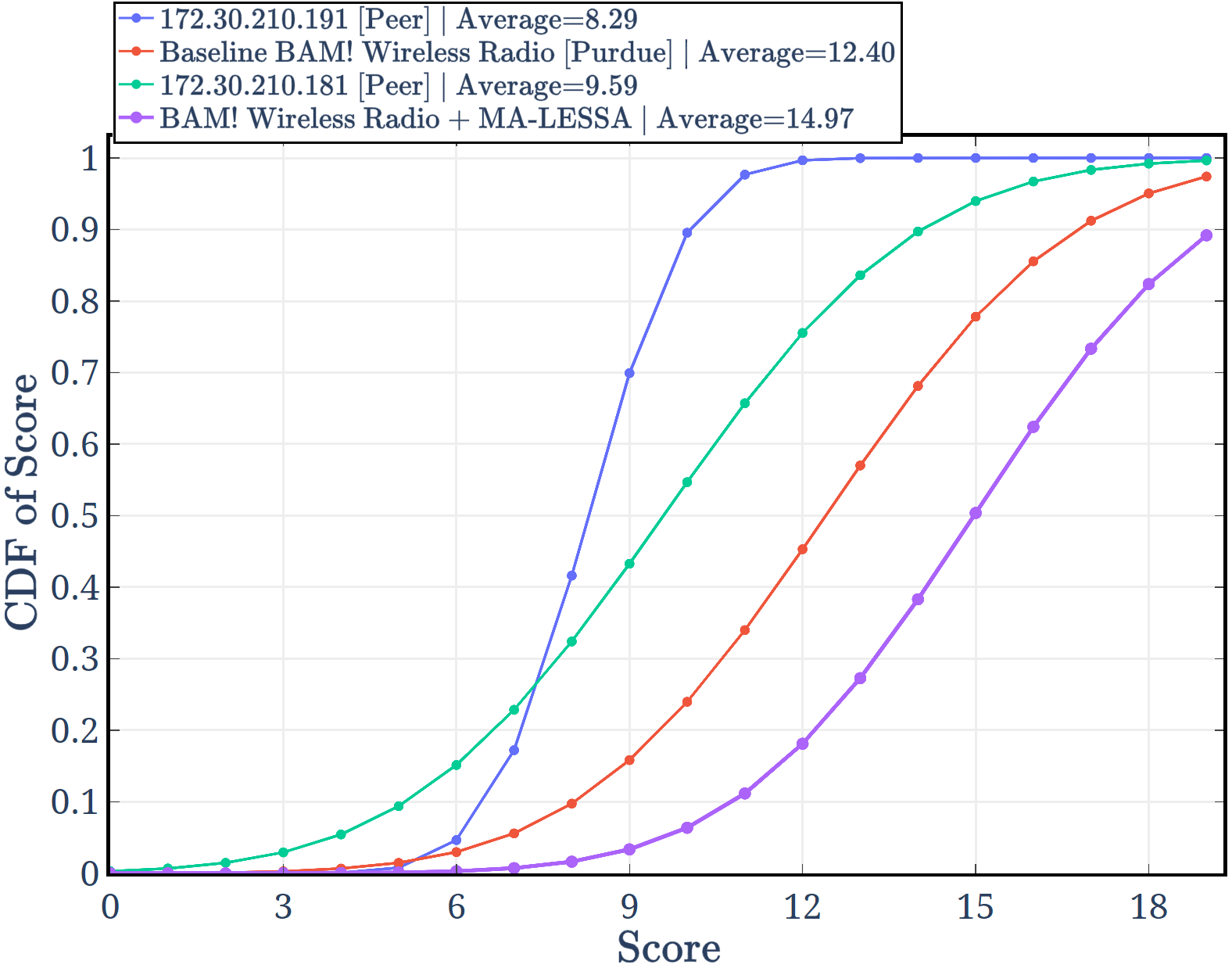}
         \caption{Centralized MA-LESSA DARPA SC$2$ Scoring Evaluation}
         \label{fig: Y. 4}
     \end{subfigure}
     \vspace{-4mm}
     \caption{CDF of the utility $R(\vec\phi(t),\vec{B}(t))$ (see \eqref{12a}) for distributed MA-LESSA against state-of-the-art (a); CDF of the Score achieved during the DARPA SC2 Active Incumbent scenario emulation: comparison between BAM! Wireless CR network design retrofitted with MA-LESSA and other competitor network radio designs (including the baseline BAM! Wireless CR network) (b).}
     \label{Fig. combo_darpa_dist}
     \vspace{-8mm}
\end{figure}

\figref{fig: Z. 2} plots the CDF of the utility achieved by
 MA-LESSA against other distributed multi-agent schemes in the state-of-the-art. We find that MA-LESSA outperforms the distributed, cooperative, $\epsilon$-greedy TD-SARSA with Linear Function Approximation framework from \cite{WCL:5} by $43$\%; it also outperforms the distributed, cooperative, time-decaying $\epsilon$-greedy algorithm with channel access rank pre-allocations from \cite{WCL:MIT} by $84$\%; and it outperforms the distributed, opportunistic, g-statistics algorithm with ACKs (without channel access rank pre-allocations) from \cite{WCL:MIT} by $324$\%. These enhancements can be attributed to the fact that MA-LESSA exploits the LUs' occupancy correlation structure across both time and frequency (vs time only of \cite{WCL:5} and the independence of \cite{WCL:MIT})
  and it leverages collaboration among CRs in the network for access order determination and sensed data aggregation.
\vspace{-4mm}

\subsection{Centralized MA-LESSA: DARPA SC$2$ Emulations}\label{Y}
In order to evaluate the performance of MA-LESSA in real-world settings, we retrofit it into the MAC layer (channel \& bandwidth allocation) of our BAM! Wireless radio \cite{BAM}, by leveraging the aggregated PSD measurements obtained at the gateway node of our BAM! Wireless network, as shown in \figref{fig:psd}. We analyze its operational capabilities in the DARPA SC2 Active Incumbent scenario \cite{DARPA:ActiveIncumbent} emulated on the Colosseum \cite{DARPA:SC2c2api}. The DARPA SC2 Active Incumbent scenario consists of a Terminal Doppler Weather Radar (TDWR) system functioning as the LU, and $5$ competitor networks (ours included), each comprising $2$ UNII WLANs: $2$ Access Points (APs) and $4$ STAtions (STAs) per AP, serving as the CRs, in a $10$MHz radio environment ($995$MHz to $1005$MHz), for $330$ seconds of emulation on the Colosseum \cite{DARPA:ActiveIncumbent}. During the Active Incumbent scenario emulation, every competitor network receives network flows from the Colosseum, which need to be delivered to the appropriate destination nodes within the network, while satisfying the imposed QoS mandates per flow (for example: max latency, min throughput, file transfer deadline, etc.). If the QoS mandates imposed on a particular network flow have been satisfied for a pre-specified period of time, then the Individual Mandates (IMs) associated with the flow are said to have been met.

Let $\mathcal{V}_{t}$ be the set of IMs achieved by a participant network in time-slot $t$, and let $\text{p}_{v}$ be the points received for satisfying an IM $v{\in}\mathcal{V}_{t}$: we define the \emph{score} of a participant network corresponding to a certain time-slot $t$ as $\sum_{v{\in}\mathcal{V}_{t}} \text{p}_{v}$ \cite{DARPA:ActiveIncumbent}. We evaluate the performance of MA-LESSA retrofitted into our standard BAM! Wireless radio \cite{BAM},
and compare it against a baseline channel and bandwidth allocation scheme that involves the weighted combination of PSD observations and collaboration data received from the other competing networks (titled "Baseline BAM! Wireless Radio [Purdue]"), and against the designs of other competitors (identified by their collaboration network registered IP address \cite{DARPA:CIL}, "172.30.210.191 [Peer]" and "172.30.210.181 [Peer]"). \figref{fig: Y. 4} plots the CDF of the score achieved: remarkably "BAM! Wireless Radio + MA-LESSA" outperforms the baseline BAM! Wireless design by $21$\% in average score, and outperforms the competitors by $56$\% and $81$\%, respectively. This result shows the potential of MA-LESSA to optimize higher-layer metrics (the network score) thanks to improved detection of white-spaces.
\begin{figure} [t]
     \begin{subfigure}{0.5\linewidth}
         \centering
         \includegraphics[width=1.0\linewidth]{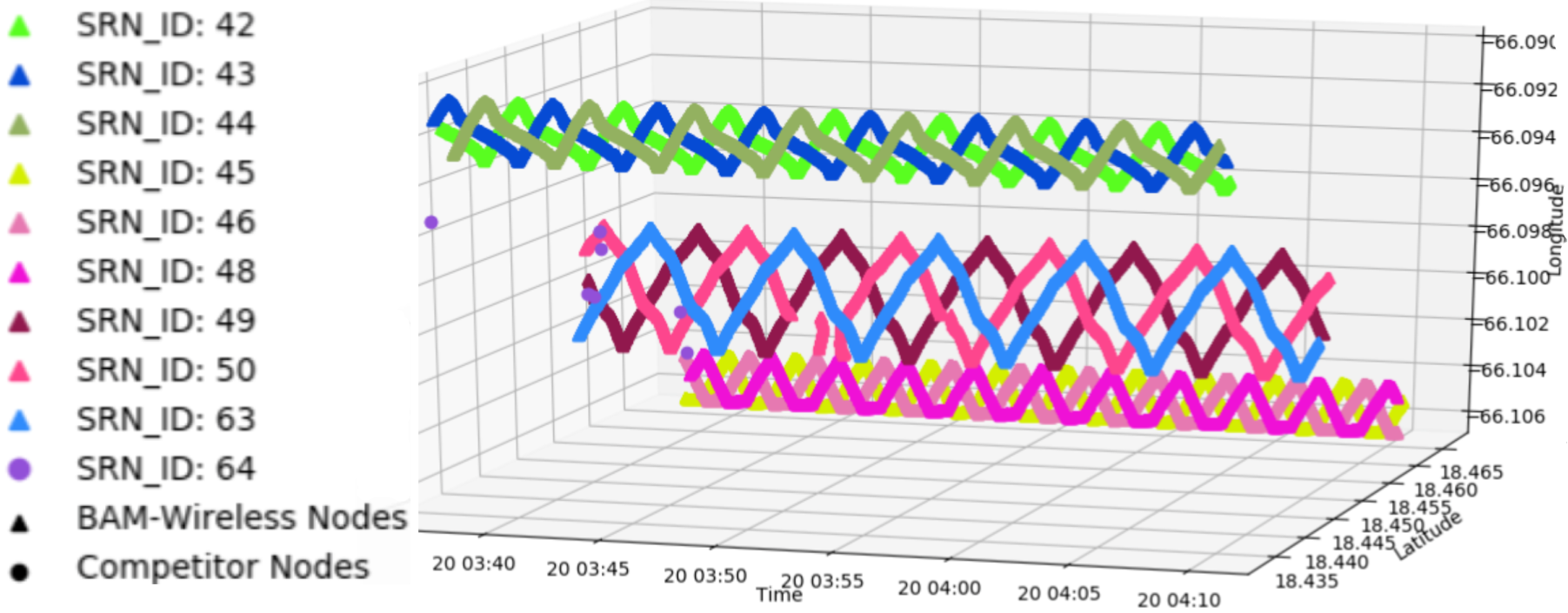}
         \caption{Payline Node Mobility}
         \label{fig:payline_gps}
     \end{subfigure}
     \begin{subfigure}{0.5\linewidth}
         \centering
         \includegraphics[width=0.8\linewidth]{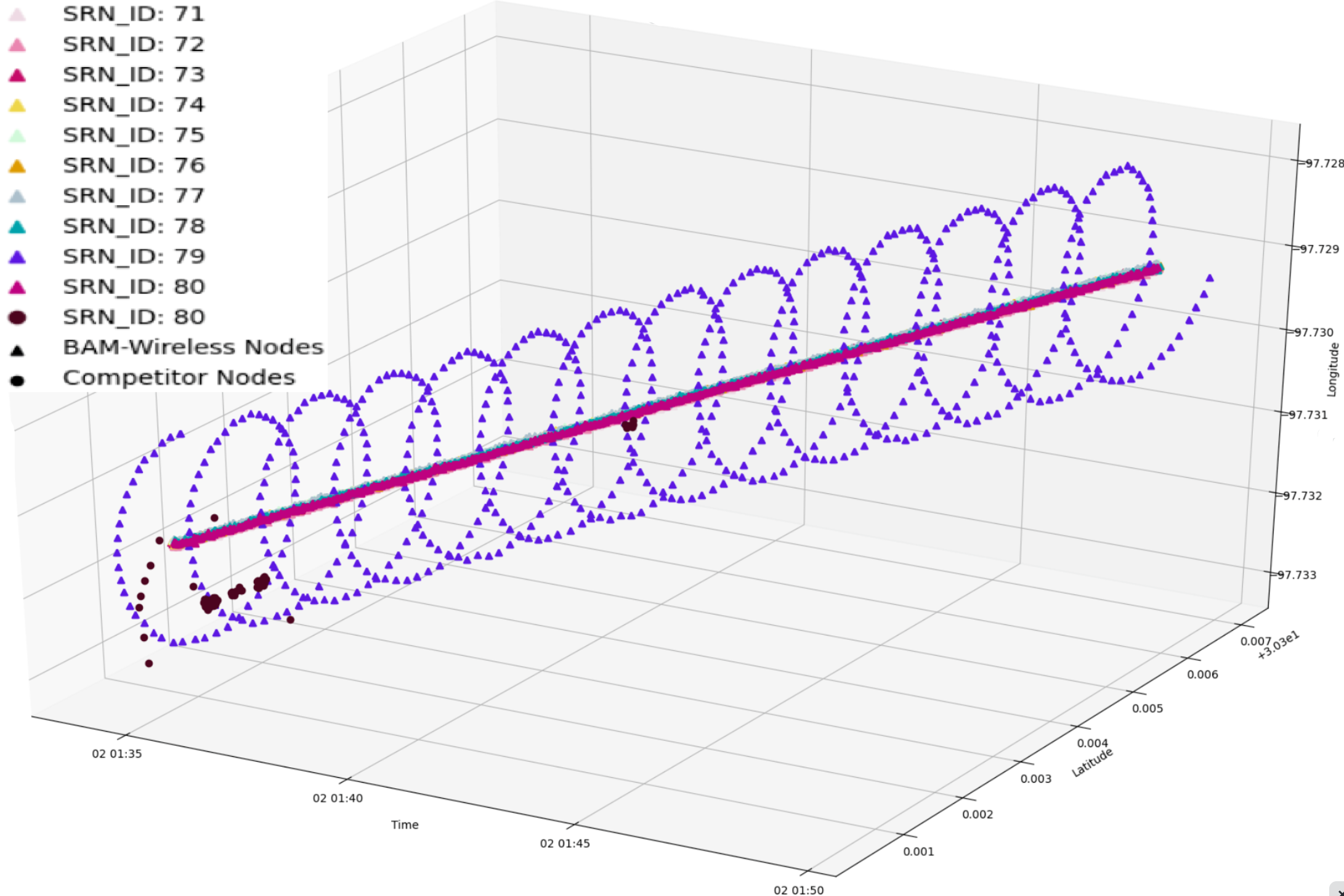}
         \caption{Alleys of Austin Node Mobility}
         \label{fig:alleys_gps}
     \end{subfigure}
     \vspace{-4mm}
     \caption{\hlt{The mobility patterns of constituent nodes in the DARPA SC$2$ Payline (a) and Alleys of Austin (b) scenarios. A Standard Radio Node (SRN) is a CR in DARPA SC2.}}
     \label{fig:gps}
     \vspace{-8mm}
\end{figure}
\label{TC:DARPASC2}
\hlt{To evaluate the proposed neighbor discovery (RSSI thresholding) and channel access rank allocation (quorum-based preferential ballot) heuristics from a computational time complexity perspective, we retrofit these schemes into the control channel design, collaboration, and data aggregation modules of the Purdue BAM! Wireless radio \cite{BAM},  and analyse their feasibility in emulations of highly mobile real-world scenarios -- namely, military deployments in the Alleys of Austin scenario \cite{DARPA:Alleys} (urban: $5\times$ $[9{-}\text{guardsmen} + 1{-}\text{UAV}]$) and disaster relief deployments in the Payline scenario \cite{DARPA:Payline} (urban: $5\times$ $[9{-}\text{EMTs} + 1{-}\text{HQ}]$). Along with the scenario-specific node mobility patterns (\figref{fig:gps}), the results of these emulations are shown in \figref{fig:nd_car_analysis}: neighbor discovery list changes are minimal in spite of node mobility (with an RSSI threshold of $22$dB), and distributed convergence of channel access rank allocation across the ensemble is achieved within the first few iterations in any given $10$s time-step.}
\begin{figure} [t]
     \begin{subfigure}{0.5\linewidth}
         \centering
         \includegraphics[width=0.9\linewidth]{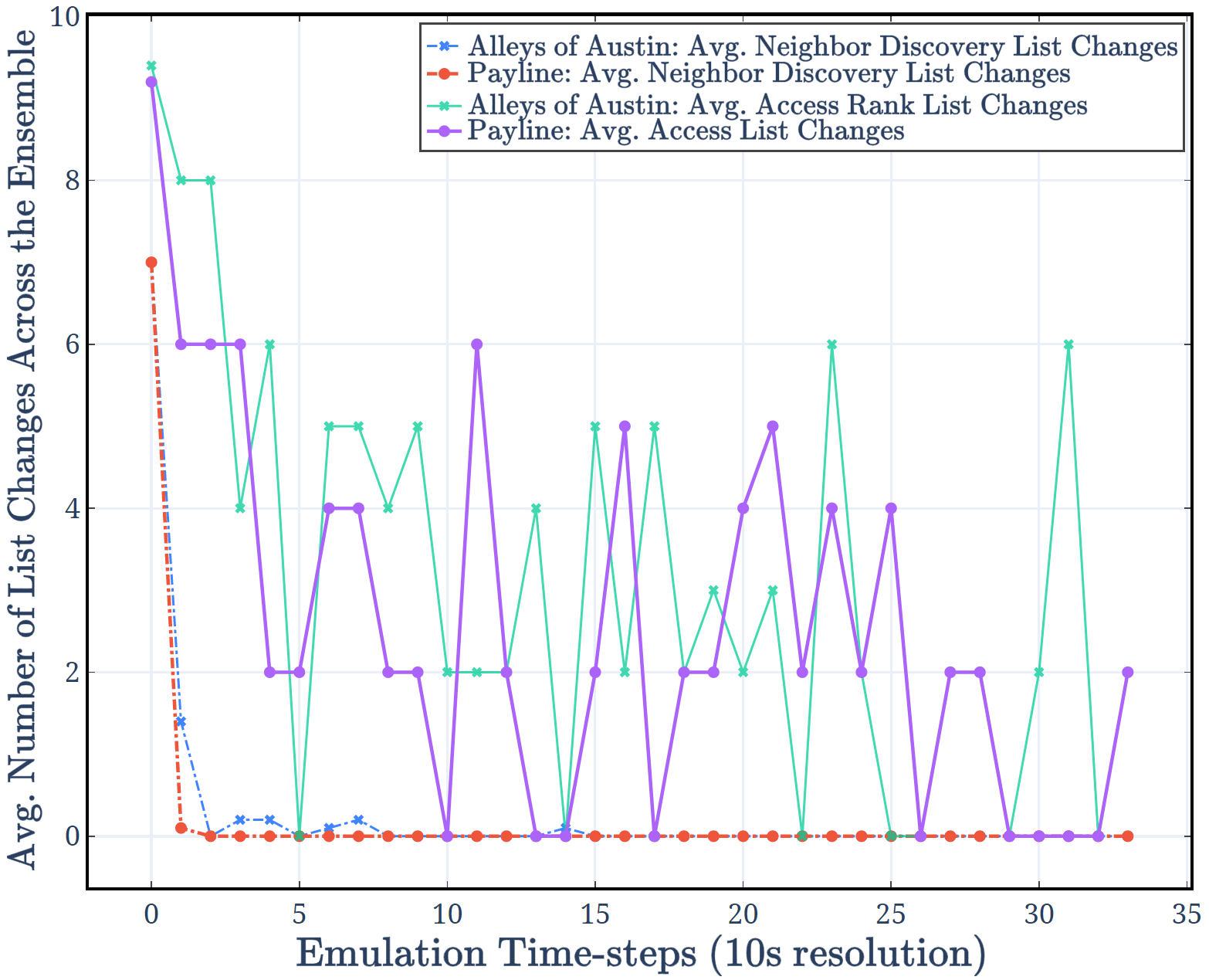}
         \caption{Neighbor Discovery and Access Rank List Changes}
         \label{fig:list_changes}
     \end{subfigure}
     \begin{subfigure}{0.5\linewidth}
         \centering
         \includegraphics[width=0.9\linewidth]{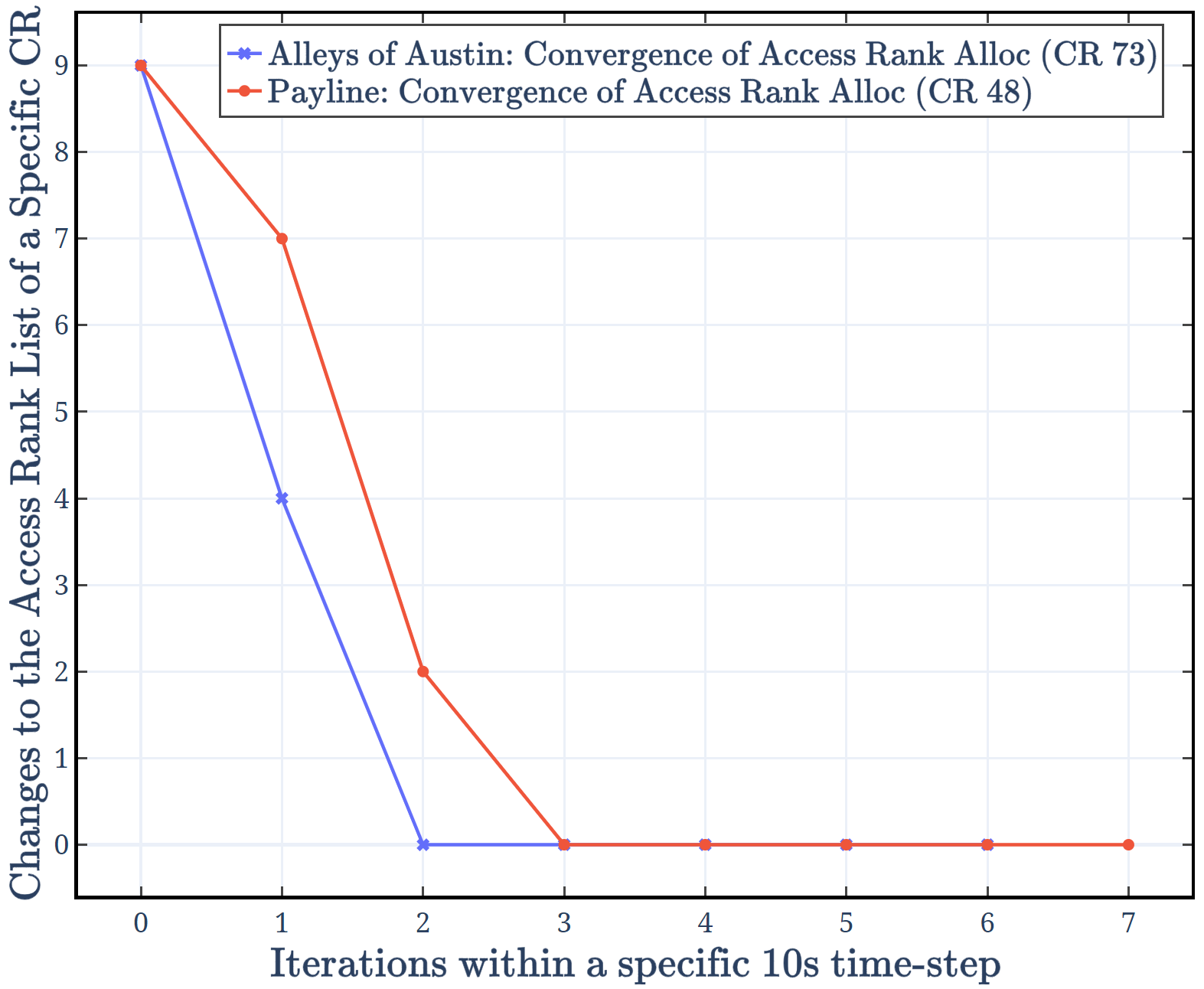}
         \caption{Convergence of Access Rank Allocation}
         \label{fig:rank_convergence}
     \end{subfigure}
     \vspace{-4mm}
     \caption{\hlt{Evaluation of neighbor discovery and channel access rank allocation: average number of neighbor discovery and channel access rank list changes across the ensemble (a); and convergence visualization of channel access rank allocation in a specific $10$s time-step for specific CRs in the network (b).}}
     \label{fig:nd_car_analysis}
     \vspace{-10mm}
\end{figure}
\vspace{-6mm}
\subsection{Implementation Feasibility of LESSA on an ESP$32$ WLAN Testbed}\label{D}
We employ $8$ ESP$32$ radios \cite{Espressif:ESP32},  each embedded on a GCTronic e-puck2 robot \cite{GCTronic:epuck2}, categorized into a network of $3$ LUs (and their $3$ corresponding sinks) occupying $6$ subcarriers in the $2.4$GHz Wi-Fi spectrum; and $2$ independent CRs, each having the capability to sense only one channel at a time. With $2$ CRs, the overall sensing restriction of the CR network is thus $\kappa=2$ (work split over $2$ ESP$32$ radios due to design limitations). The 3 LUs occupy the spectrum according to a Markovian time-frequency correlation structure (described by \eqref{6}), with parameters $p_{0,0}{=}0.1$, $p_{0,1}{=}p_{1,0}{=}0.3$, $p_{1,1}{=}0.7$ and $q_{0}{=}0.3$, $q_{1}{=}0.8$.
We simulate the occupancy behavior of the LUs according to
this time-frequency correlation structure, and solve for the spectrum sensing and access policy using LESSA, by mimicking the observational capabilities of the actual ESP$32$ radios. Note that this step is performed on a PC.

The simulated LU occupancy behavior and the time-slot specific channel access decisions (derived through LESSA), are stored in databases for export onto the ESP$32$ network. In time-slot $t$, a peer-to-peer communication link $l_{ij}$ is established between an ESP$32$ LU $j{\in}\{1,2,3\}$ and its designated sink $i{\in}\{1,2,3\}$, over a subcarrier $k_{l_{ij}} = k{\in}\{1,2,{\dots},6\}$, as determined by the exported LU occupancy database. These channel allocations are done orthogonally at the LUs, so that $k_{l_{ij}}{\neq}k_{l_{i',j'}},{\forall}i,i'{\in}\{1,2,3\},$ $j,j'{\in}\{1,2,3\}$. In the next synchronized time-slot $t+1$, this link $l_{ij}$ moves to channel $k'{\in}\{1,2,{\dots},6\}$, again as determined by the LU's behavioral rules maintained in the exported occupancy database. This same procedure takes place for the other two LU communication links in every time-slot until the end of the evaluation period.

Although the PC-based POMDP solver employs a single CR which can access $2$ subcarriers at a time in order to deliver its flows (see the access part of the POMDP formulation in \secref{II.0}), we employ $2$ ESP$32$ CR radios in the network (serving as one), with the channel access work synchronously and evenly split between the two, due to the actual physical design limitations of the ESP$32$ radio, each able to access only one channel at a time. More specifically, we split the $2$-channel access decision in time-slot $t$, as determined by the time-slot specific POMDP channel access database, into a $1$-channel access action at each ESP$32$ CR radio. Next, based on whether the channel access at the $2$ ESP$32$ CR radios was successful, we compute the success rate as $\frac{1}{2}\sum_{j=1}^{2}\mathcal{I}[B_{k_{CR_{j}}}(t)=0]$,
where $B_{k_{CR_{j}}}(t){\in}\{0,1\}$ is the occupancy variable of the channel accessed by CR $j$ in time-slot $t$. The implementation of LESSA on this ad-hoc WLAN testbed demonstrates a channel access success probability of $96$\% across an evaluation period of $300$s.
\vspace{-4mm}
\section{Conclusion}\label{V}
In this paper, we formulate a learning-based spectrum sensing and access problem in resource-constrained radio ecosystems as an approximate POMDP, which leverages learning of the LU occupancy correlation model via the Baum-Welch algorithm and solving for an approximately optimal sensing and access policy via the PERSEUS algorithm. We propose fragmentation, Hamming distance state filter heuristics, and Monte-Carlo methods to alleviate the inherent computational complexity of PERSEUS. Through system simulations, we demonstrate the advantages of exploiting the correlation structure -- as opposed to Neyman-Pearson Detection which assumes independence; adapting the spectrum sensing decision to enhance white-space detection -- as opposed to Viterbi, which uses a fixed sensing strategy; and a parametric learning and adaptation framework -- over model-free Deep Q-Network designs. We also demonstrate the feasibility of a concurrent learning and decision-making framework, as opposed to state-of-the-art correlation-coefficient based clustering algorithms, which rely on pre-loaded datasets for determining the correlation in the LU occupancy behavior. Our framework enables a critical feature in practical scenarios: the ability of the CR to regulate the interference caused to LUs, by adjusting a penalty parameter. Also, extending our single-agent model to multi-agent settings, we demonstrate superior performance over the state-of-the-art, in both centralized and distributed deployment settings (collaborative and opportunistic access).
\vspace{-4mm}
\bibliographystyle{IEEEtran}
\bibliography{main}

\begin{thebibliography}{10}
\providecommand{\url}[1]{#1}
\csname url@samestyle\endcsname
\providecommand{\newblock}{\relax}
\providecommand{\bibinfo}[2]{#2}
\providecommand{\BIBentrySTDinterwordspacing}{\spaceskip=0pt\relax}
\providecommand{\BIBentryALTinterwordstretchfactor}{4}
\providecommand{\BIBentryALTinterwordspacing}{\spaceskip=\fontdimen2\font plus
\BIBentryALTinterwordstretchfactor\fontdimen3\font minus
  \fontdimen4\font\relax}
\providecommand{\BIBforeignlanguage}[2]{{%
\expandafter\ifx\csname l@#1\endcsname\relax
\typeout{** WARNING: IEEEtran.bst: No hyphenation pattern has been}%
\typeout{** loaded for the language `#1'. Using the pattern for}%
\typeout{** the default language instead.}%
\else
\language=\csname l@#1\endcsname
\fi
#2}}
\providecommand{\BIBdecl}{\relax}
\BIBdecl

\bibitem{ICC:paper}
B.~{Keshavamurthy} and N.~{Michelusi}, ``{Learning-based Cognitive Radio Access
  via Randomized Point-Based Approximate POMDPs},'' 2020, {Presented at IEEE
  ICC 2021}.

\bibitem{5GSurvey}
M.~Shafi \emph{et~al.}, ``5g: A tutorial overview of standards, trials,
  challenges, deployment, and practice,'' \emph{IEEE Journal on Selected Areas
  in Communications}, vol.~35, no.~6, pp. 1201--1221, 2017.

\bibitem{WCL:3}
S.~{Maleki}, S.~P. {Chepuri}, and G.~{Leus}, ``{Energy and throughput efficient
  strategies for cooperative spectrum sensing in cognitive radios},'' in
  \emph{IEEE 12th International Workshop on Signal Processing Advances in
  Wireless Communications}, June 2011, pp. 71--75.

\bibitem{WCL:4}
K.~{Cohen}, Q.~{Zhao}, and A.~{Scaglione}, ``{Restless Multi-Armed Bandits
  under time-varying activation constraints for dynamic spectrum access},'' in
  \emph{48th Asilomar Conference on Signals, Systems and Computers}, Nov 2014,
  pp. 1575--1578.

\bibitem{WCL:10}
N.~{Michelusi}, M.~{Nokleby}, U.~{Mitra}, and R.~{Calderbank}, ``{Multi-Scale
  Spectrum Sensing in Dense Multi-Cell Cognitive Networks},'' \emph{IEEE
  Transactions on Communications}, vol.~67, no.~4, pp. 2673--2688, April 2019.

\bibitem{WCL:11}
S.~{Mosleh}, A.~A. {Tadaion}, and M.~{Derakhtian}, ``{Performance analysis of
  the Neyman-Pearson fusion center for spectrum sensing in a Cognitive Radio
  network},'' in \emph{IEEE EUROCON}, May 2009, pp. 1420--1425.

\bibitem{WCL:5}
J.~{Lund\'{e}n}, S.~R. {Kulkarni}, V.~{Koivunen}, and H.~V. {Poor},
  ``{Multiagent Reinforcement Learning Based Spectrum Sensing Policies for
  Cognitive Radio Networks},'' \emph{IEEE Journal of Selected Topics in Signal
  Processing}, vol.~7, no.~5, pp. 858--868, Oct 2013.

\bibitem{WCL:6}
C.~{Park}, S.~{Kim}, S.~{Lim}, and M.~{Song}, ``{HMM Based Channel Status
  Predictor for Cognitive Radio},'' in \emph{Asia-Pacific Microwave
  Conference}, Dec 2007, pp. 1--4.

\bibitem{WCL:8}
L.~{Ferrari}, Q.~{Zhao}, and A.~{Scaglione}, ``{Utility Maximizing Sequential
  Sensing Over a Finite Horizon},'' \emph{IEEE Transactions on Signal
  Processing}, vol.~65, no.~13, pp. 3430--3445, July 2017.

\bibitem{WCL:9}
N.~{Michelusi} and U.~{Mitra}, ``{Cross-Layer Estimation and Control for
  Cognitive Radio: Exploiting Sparse Network Dynamics},'' \emph{IEEE
  Transactions on Cognitive Communications and Networking}, vol.~1, no.~1, pp.
  128--145, March 2015.

\bibitem{WCL:12}
S.~{Yin}, D.~{Chen}, Q.~{Zhang}, M.~{Liu}, and S.~{Li}, ``{Mining Spectrum
  Usage Data: A Large-Scale Spectrum Measurement Study},'' \emph{IEEE
  Transactions on Mobile Computing}, vol.~11, no.~6, pp. 1033--1046, June 2012.

\bibitem{WCL:7}
M.~{Gao}, X.~{Yan}, Y.~{Zhang}, C.~{Liu}, Y.~{Zhang}, and Z.~{Feng}, ``{Fast
  Spectrum Sensing: A Combination of Channel Correlation and Markov Model},''
  in \emph{IEEE Military Communications Conference}, Oct 2014, pp. 405--410.

\bibitem{WCL:DQN}
S.~{Wang}, H.~{Liu}, P.~H. {Gomes}, and B.~{Krishnamachari}, ``{Deep
  Reinforcement Learning for Dynamic Multichannel Access in Wireless
  Networks},'' \emph{IEEE Transactions on Cognitive Communications and
  Networking}, vol.~4, no.~2, pp. 257--265, 2018.

\bibitem{DARPA:ActiveIncumbent}
\BIBentryALTinterwordspacing
DARPA-SC2-Freshdesk, ``{Active Incumbent Scenario Specifications},''
  \emph{DARPA Spectrum Collaboration Challenge (SC2)}, 2019. [Online].
  Available:
  \url{https://sc2colosseum.freshdesk.com/support/solutions/articles/22000239489-active-incumbent-}
\BIBentrySTDinterwordspacing

\bibitem{WCL:MIT}
A.~Anandkumar, N.~Michael, and A.~Tang, ``{Opportunistic Spectrum Access with
  Multiple Users: Learning under Competition},'' in \emph{Proceedings of the
  29th Conference on Information Communications}, ser. INFOCOM'10.\hskip 1em
  plus 0.5em minus 0.4em\relax IEEE Press, 2010, p. 803–811.

\bibitem{LSTM}
L.~Yu, Q.~Wang, Y.~Guo, and P.~Li, ``Spectrum availability prediction in
  cognitive aerospace communications: A deep learning perspective,'' in
  \emph{Cognitive Communications for Aerospace Applications Workshop (CCAA)},
  2017, pp. 1--4.

\bibitem{CNN}
B.~M. Pati, M.~Kaneko, and A.~Taparugssanagorn, ``A deep convolutional neural
  network based transfer learning method for non-cooperative spectrum
  sensing,'' \emph{IEEE Access}, vol.~8, pp. 164\,529--164\,545, 2020.

\bibitem{DARPA:SC2}
\BIBentryALTinterwordspacing
M.~Rosker, ``{Spectrum Collaboration Challenge (SC2)},'' \emph{DARPA Spectrum
  Collaboration Challenge (SC2)}, 2018. [Online]. Available:
  \url{https://www.darpa.mil/program/spectrum-collaboration-challenge}
\BIBentrySTDinterwordspacing

\bibitem{Espressif:ESP32}
\BIBentryALTinterwordspacing
{Espressif Systems (Shanghai) Co. Ltd.}, ``{Espressif ESP32: A Different IoT
  Power and Performance},'' \emph{Espressif}, 2019. [Online]. Available:
  \url{https://www.espressif.com/en/products/hardware/esp32/overview}
\BIBentrySTDinterwordspacing

\bibitem{Rabiner_1989}
L.~R. Rabiner, ``{A Tutorial on Hidden Markov Models and Selected Applications
  in Speech Recognition},'' \emph{Proceedings of the IEEE 77, no. 2}, pp.
  257--286, Feb 1989.

\bibitem{WCL:13}
\BIBentryALTinterwordspacing
M.~T.~J. Spaan and N.~A. Vlassis, ``{Perseus: Randomized Point-based Value
  Iteration for POMDPs},'' \emph{CoRR}, vol. abs/1109.2145, 2011. [Online].
  Available: \url{http://arxiv.org/abs/1109.2145}
\BIBentrySTDinterwordspacing

\bibitem{MDND}
H.~Park, Y.~Kim, T.~Song, and S.~Pack, ``Multiband directional neighbor
  discovery in self-organized mmwave ad hoc networks,'' \emph{IEEE Transactions
  on Vehicular Technology}, vol.~64, no.~3, pp. 1143--1155, 2015.

\bibitem{AMQ}
\BIBentryALTinterwordspacing
{Apache ActiveMQ}, ``{Clusters: ActiveMQ Artemis Documentation},''
  \emph{ActiveMQ Artemis Documentation}, 2014. [Online]. Available:
  \url{https://activemq.apache.org/components/artemis/documentation/latest/clusters.html}
\BIBentrySTDinterwordspacing

\bibitem{GCTronic:epuck2}
\BIBentryALTinterwordspacing
{GCTronic}, ``{Epuck2 Specifications and General Wiki},'' \emph{{GCTronic}
  e-puck2 online wiki}, 2020. [Online]. Available:
  \url{https://www.gctronic.com/doc/index.php/e-puck2}
\BIBentrySTDinterwordspacing

\bibitem{BAM}
\BIBentryALTinterwordspacing
{DARPA}, ``{Purdue University BAM! Wireless Radio},'' \emph{DARPA Spectrum
  Collaboration Challenge (SC2)}, 2019. [Online]. Available:
  \url{https://archive.darpa.mil/sc2/news/spectrum-collaboration-challenge-awards-four-teams-with-half-prizes}
\BIBentrySTDinterwordspacing

\bibitem{DARPA:CIL}
\BIBentryALTinterwordspacing
{DARPA SC2 GitLab CIL Schematics}, ``{CIL Specifications},'' \emph{DARPA
  Spectrum Collaboration Challenge (SC2)}, 2019. [Online]. Available:
  \url{https://gitlab.com/darpa-sc2-phase3/CIL}
\BIBentrySTDinterwordspacing

\bibitem{DARPASC2:end1}
F.~A.~P. d.~{Figueiredo}, D.~{Stojadinovic}, P.~{Maddala}, R.~{Mennes},
  I.~{Jabandžić}, X.~{Jiao}, and I.~{Moerman}, ``{SCATTER PHY: A Physical
  Layer for the DARPA Spectrum Collaboration Challenge},'' in \emph{IEEE
  International Symposium on Dynamic Spectrum Access Networks (DySPAN)}, 2019,
  pp. 1--6.

\bibitem{8935729}
R.~J. {Baxley} and R.~S. {Thompson}, ``{Team Zylinium DARPA Spectrum
  Collaboration Challenge Radio Design and Implementation},'' in \emph{IEEE
  International Symposium on Dynamic Spectrum Access Networks (DySPAN)}, 2019,
  pp. 1--6.

\bibitem{DARPASC2:end3}
D.~{Stojadinovic}, F.~A.~P. {de Figueiredo}, P.~{Maddala}, I.~{Seskar}, and
  W.~{Trappe}, ``{SC2 CIL: Evaluating the Spectrum Voxel Announcement
  Benefits},'' in \emph{IEEE International Symposium on Dynamic Spectrum Access
  Networks (DySPAN)}, 2019, pp. 1--6.

\bibitem{8935774}
S.~{Giannoulis} \emph{et~al.}, ``{Dynamic and Collaborative Spectrum Sharing:
  The SCATTER Approach},'' in \emph{IEEE International Symposium on Dynamic
  Spectrum Access Networks (DySPAN)}, 2019, pp. 1--6.

\bibitem{DARPA:Payline}
\BIBentryALTinterwordspacing
DARPA-SC2-Freshdesk-PE2, ``{Payline (2 Stage)},'' \emph{DARPA Spectrum
  Collaboration Challenge (SC2)}, 2019. [Online]. Available:
  \url{https://sc2colosseum.freshdesk.com/support/solutions/articles/22000234543-payline-2-stage-7065-}
\BIBentrySTDinterwordspacing

\bibitem{DARPA:Alleys}
\BIBentryALTinterwordspacing
DARPA-SC2-Freshdesk-CE, ``{Alleys of Austin with Points - 5 Teams},''
  \emph{DARPA Spectrum Collaboration Challenge (SC2)}, 2019. [Online].
  Available:
  \url{https://sc2colosseum.freshdesk.com/support/solutions/articles/22000237879-pe2-alleys-of-austin-w-points-5-team-7013-}
\BIBentrySTDinterwordspacing

\bibitem{4213046}
R.~I.~C. {Chiang}, G.~B. {Rowe}, and K.~W. {Sowerby}, ``{A Quantitative
  Analysis of Spectral Occupancy Measurements for Cognitive Radio},'' in
  \emph{IEEE 65th Vehicular Technology Conference - VTC2007-Spring}, April
  2007, pp. 3016--3020.

\bibitem{McHenry:2006:CSO:1234388.1234389}
M.~A. McHenry, P.~A. Tenhula, D.~McCloskey, D.~A. Roberson, and C.~S. Hood,
  ``{Chicago Spectrum Occupancy Measurements \& Analysis and a Long-term
  Studies Proposal},'' in \emph{Proceedings of the First International Workshop
  on Technology and Policy for Accessing Spectrum}, ser. TAPAS '06.\hskip 1em
  plus 0.5em minus 0.4em\relax New York, NY, USA: ACM, 2006.

\bibitem{BIC}
D.~M.~A. Haughton, ``On the choice of a model to fit data from an exponential
  family,'' \emph{The Annals of Statistics}, vol.~16, no.~1, pp. 342--355,
  1988.

\bibitem{PUOccupancy:18}
L.~P. Kaelbling, M.~L. Littman, and A.~R. Cassandra, ``{Planning and Acting in
  Partially Observable Stochastic Domains},'' \emph{Artif. Intell.}, vol. 101,
  no. 1–2, p. 99–134, May 1998.

\bibitem{Baum_1966}
L.~E. Baum and T.~Petrie, ``Statistical inference for probabilistic functions
  of finite state markov chains,'' \emph{Ann. Math. Statist.}, vol.~37, no.~6,
  pp. 1554--1563, 12 1966.

\bibitem{PUOccupancy:17}
J.~Pineau, G.~Gordon, and S.~Thrun, ``{Point-Based Value Iteration: An Anytime
  Algorithm for POMDPs},'' in \emph{Proceedings of the 18th International Joint
  Conference on Artificial Intelligence}, ser. IJCAI’03.\hskip 1em plus 0.5em
  minus 0.4em\relax San Francisco, CA, USA: Morgan Kaufmann Publishers Inc.,
  2003, p. 1025–1030.

\bibitem{Tse}
D.~Tse and P.~Viswanath, \emph{{Fundamentals of Wireless Communication}}.\hskip
  1em plus 0.5em minus 0.4em\relax USA: Cambridge University Press, 2005.

\bibitem{Urban_Model}
Q.~Feng, J.~McGeehan, E.~Tameh, and A.~Nix, ``Path loss models for
  air-to-ground radio channels in urban environments,'' in \emph{2006 IEEE 63rd
  Vehicular Technology Conference}, vol.~6, 2006, pp. 2901--2905.

\bibitem{Convex}
Y.~Sun, A.~Baricz, and S.~Zhou, ``On the monotonicity, log-concavity, and tight
  bounds of the generalized marcum and nuttall $q$-functions,'' \emph{IEEE
  Transactions on Information Theory}, vol.~56, no.~3, pp. 1166--1186, 2010.

\bibitem{Duplyakin+:ATC19}
D.~Duplyakin \emph{et~al.}, ``The design and operation of {CloudLab},'' in
  \emph{Proceedings of the {USENIX} Annual Technical Conference (ATC)}, Jul.
  2019, pp. 1--14.

\bibitem{DARPA:SC2c2api}
\BIBentryALTinterwordspacing
DARPA, ``{Radio Command and Control API},'' \emph{DARPA Spectrum Collaboration
  Challenge (SC2)}, 2018. [Online]. Available:
  \url{https://sc2colosseum.freshdesk.com/support/solutions/articles/22000220460-radio-command-and-control-c2-api}
\BIBentrySTDinterwordspacing

\end{thebibliography}
\end{document}